\documentclass[a4paper, traditabstract,usenames,dvipsnames]{aa} 
\pdfoutput=1
\usepackage{txfonts}
\usepackage{bbm}
\usepackage{microtype}
\usepackage{url}
\usepackage{booktabs}
\usepackage{commath}
\usepackage[bookmarks, colorlinks, breaklinks, ]{hyperref}  
\hypersetup{linkcolor=blue,citecolor=MidnightBlue,filecolor=black,urlcolor=MidnightBlue} 
\newfont{\gwpfont}{cmssq8 scaled 1000}
\newcommand{\rexcess}{{\gwpfont REXCESS}}

\usepackage{subcaption}
\usepackage[labelsep=quad,skip=5pt,labelfont=bf]{caption}
\def\msol{{M$_{\odot}$}}

\def\xmm{XMM-{\it Newton}}
\def\planck{{\it Planck}}
\def\chandra{{\it Chandra}}

\def\M500{M_{500}}
\def\R500{R_{500}}

\def\YX {Y_{\rm X, 500}}
\def\TX {T_{\rm X}}

\def\fgv {f_{\rm g,500}}

\def\Mv {M_{\rm 500}}
\def \Rv {R_{500}}
\def\keV {\rm keV}

\def\Lx{$L_{\rm X}$}

\def\Lxc{$L_{\rm Xc}$}
\def\Lxcp{L_{\rm Xc}^{\prime}}

\def \fgv  {f_{\rm gas, 500}}

\def \Mgas  {M_{\rm gas}}

\def \Mv  {M_{\rm 500}}
\def \RV  {R_{\rm V}}
\def \MV  {M_{\rm V}}
\def \Rv  {R_{500}}

\def \rhoc {\rho_{\rm c}}

\def \rhocz {\rhoc(z)}
\def \rhov{\rho_{\rm 500}}
\def \rhog {\rho_{\rm gas}}
\def \rhom{\rho_{\rm m}}

\def \zi {z_{\rm i}}
\def \Mvi {M_{\rm 500,i}}

\def \xij {x_{\rm i,j}}
\def \rhoij {\rho_{\rm i,j}}
\def \rhomij {\rho_{\rm m,i,j}}
\def \sigdij {\sigma^2_{\rm i,j}}
\def \aM {\alpha_{\rm M}}
\def \az  {\alpha_{\rm z}}
\def \xs{x_{\rm s}}

\def\msol {{\rm M_{\odot}}}

\def\lesssim{\mathrel{\hbox{\rlap{\hbox{\lower4pt\hbox{$\sim$}}}\hbox{$<$}}}}
\def\gtrsim{\mathrel{\hbox{\rlap{\hbox{\lower4pt\hbox{$\sim$}}}\hbox{$>$}}}}

\newcommand{\propsim}{\lower 3pt \hbox{$\, \buildrel {\textstyle
       \propto}\over {\textstyle \sim}\,$}}

\begin{document}
\title{Linking a universal gas density profile to the core-excised X-ray luminosity in galaxy clusters up to $z\sim1.1$}   
\author{G.W. Pratt\inst{1} \and M. Arnaud\inst{1} \and B.J. Maughan\inst{2} \and J.-B. Melin\inst{3}
}
\authorrunning{G.W. Pratt et al.}
\titlerunning{The universal galaxy cluster gas density profile and the \Lxc$-M$ relation}
 \institute{
 $^1$ Université Paris-Saclay, Université Paris Cité, CEA, CNRS, AIM de Paris-Saclay, 91191 Gif-sur-Yvette, France \\
 \email{gabriel.pratt@cea.fr} \\ 
 $^2$ HH Wills Physics Laboratory, University of Bristol, Tyndall Ave, Bristol, BS8 1TL, UK \\
  $^3$ IRFU, CEA, Université Paris-Saclay, 91191 Gif-sur-Yvette, France
}
\date{Received 10 January 2022; accepted 26 May 2022}
\abstract{
We investigate the regularity of galaxy cluster gas density profiles and the link to the relation between core-excised luminosity, \Lxc, and mass from the $Y_{\rm X}$ proxy, $M_{\rm Y_{X}}$, for 93 objects selected through their Sunyaev-Zeldovich effect (SZE) signal. The sample spans a mass range of $\Mv = [0.5 - 20] \times 10^{14}$ M$_{\odot}$, and lies at redshifts $0.05 < z < 1.13$. To investigate differences in X-ray and SZE selection, we compare to the local X-ray-selected \rexcess\ sample. Using XMM-{\it Newton} observations, we derive an average intra-cluster medium (ICM) density profile for the SZE-selected systems and determine its scaling with mass and redshift. This average profile exhibits an evolution that is slightly stronger than self-similar ($\alpha_{\rm z} = 2.09\pm 0.02$), and a significant dependence on mass ($\alpha_{\rm M} = 0.22\pm0.01$). Deviations from this average scaling with radius, which we quantify, indicate different evolution for the core regions as compared to the bulk. We measure the radial variation of the intrinsic scatter in scaled density profiles, finding a minimum of $\sim 20\%$ at $R\sim [0.5-0.7]\,\Rv$ and a value of $\sim 40\%$ at $\Rv$; moreover, the scatter evolves slightly with redshift. The average profile of the SZE-selected systems adequately describes the X-ray-selected systems and their intrinsic scatter at low redshift, except in the very central regions. We examine the evolution of the scaled core properties over time, which are positively skewed at later times, suggesting an increased incidence of centrally peaked objects at lower redshifts. The relation between core-excised luminosity, \Lxc, and mass is extremely tight, with a measured logarithmic intrinsic scatter of $\sigma_{\ln{L_{\rm Xc}|M_{\rm Yx}}} \sim 0.13$. Using extensive simulations, we investigate the impact of selection effects, intrinsic scatter, and covariance between quantities on this relation. The slope is insensitive to selection and intrinsic scatter between quantities; however, the scatter is very dependent on the covariance between \Lxc\ and $Y_{\rm X}$.  Accounting for our use of the $Y_{\rm X}$ proxy to determine the mass, for observationally motivated values of covariance we estimate an upper limit to the logarithmic intrinsic scatter with respect to the true mass of $\sigma_{\ln{L_{\rm Xc}|M}} \sim 0.22$. We explicitly illustrate the connection between the scatter in density profiles and that in the \Lxc$-M$ relation. Our results are consistent with the overall conclusion that the ICM bulk evolves approximately self-similarly, with the core regions evolving separately. They indicate a systematic variation of the gas content with mass. They also suggest that the core-excised X-ray luminosity, \Lxc, has a tight and well-understood relation to the underlying mass.
}


\maketitle
%

\section{Introduction}\label{sec:intro}

In a $\Lambda$ cold dark matter Universe, halo assembly is driven by the hierarchical  gravitational collapse of the dominant dark matter component. To first order, this process is self-similar and scale-free because gravity has no characteristic scale.  If the baryon content remains constant, power-law relations link the baryonic observable properties -- such as X-ray luminosity \Lx, or Sunyaev-Zeldovich effect (SZE) signal $Y_{\rm SZ}$, or total optical richness $\Lambda$ -- to the cluster mass and redshift. Each of these observables is sensitive to a different underlying  intrinsic physical characteristic (e.g. the distribution of gas or the number of red sequence galaxies above a given threshold). However, the detection of a baryonic observable also depends on the intrinsic properties of the signal itself. For example, while the SZE signal is proportional to the gas density, the X-ray emission is proportional to the square of the gas density. This means that X-ray measurements are very sensitive to the physical conditions in the core regions, while SZE measurements are much less so.

The X-ray luminosity, \Lx, is an attractive quantity because it can be measured from very few source counts once the redshift is known. For a virialised galaxy cluster where the intra-cluster medium (ICM) is in hydrostatic equilibrium, \Lx\ depends only on the halo mass, $M$, redshift, $z$, and the distribution of the gas in the dark matter potential. Power-law $L_{\rm X}-M$ relations have indeed been observed \citep[e.g.][]{mau07,ryk08,zha08,con14,pra09,vik09,sr17a,lov20}. However,  these relations exhibit a large intrinsic scatter ($\sim 40$ per cent), linked to the presence of cool cores and merging activity \citep[e.g.][]{pra09}. Exclusion of the core regions, by measuring \Lx\ in an annulus excluding the cluster centre, significantly reduces the intrinsic scatter \citep{fab94,mau07,pra09,man18}.

%
%
\begin{figure*}[!ht]
\begin{center}
\includegraphics[bb=35 605 560 790 ,clip,scale=1.0,angle=0,keepaspectratio,width=1.\textwidth]{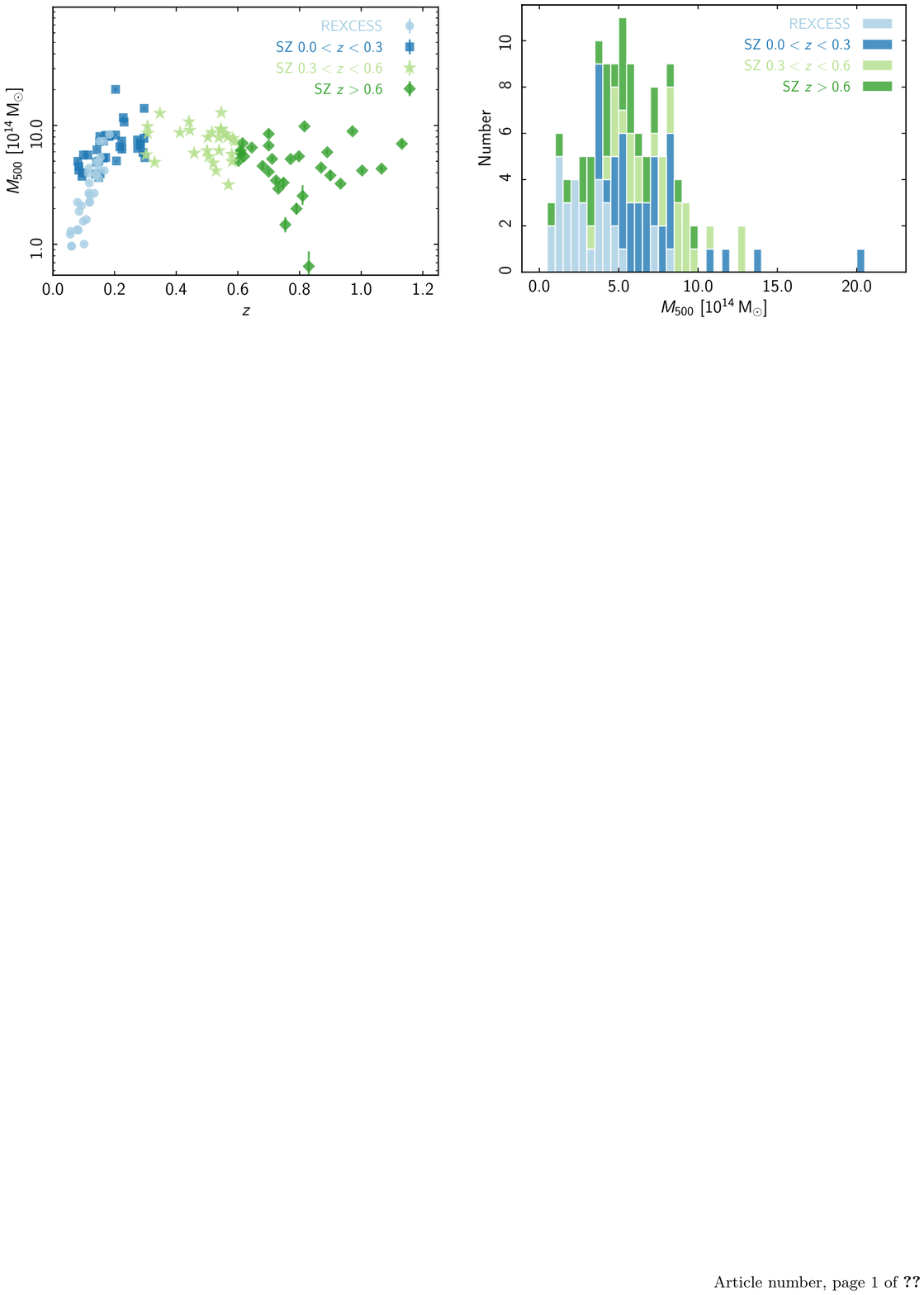}
\end{center}
\caption{\footnotesize Sample properties. {\it Left:} Redshift-mass distribution of the clusters used in this paper. The SZE-selected clusters comprise a subset of 44 systems from the \planck\ Early SZ sample \citep{esz} at $z<0.5$ and a further 49 clusters at $z>0.5$. \rexcess\ \citep{boe07} is an X-ray-selected sample of 31 objects at $z<0.25$. {\it Right:} Stacked histogram of the mass distribution. The \rexcess\ sample has a lower median mass than the SZE-selected samples.
}
\label{fig:sample}
\end{figure*}
%

Radial gas density profiles have been a key means of obtaining information about the ICM since the advent of X-ray imaging. The high spatial resolution observations afforded by \xmm\ and \chandra\ observations have revealed the complexity of both the core regions, which are strongly affected by non-gravitational processes \citep{cro08, pra10}, and the outskirts, where the gas distribution becomes progressively more inhomogeneous \citep[e.g.][]{eck15}. Key open questions are how the ICM evolves over time in the dark matter potential, and how this connects to the formation and evolution of cool cores.

The advent of SZE surveys has resulted in the detection of large numbers of clusters at $z>0.5$ \citep{has13, ble15,psz2,hil21}, extending the redshift leverage for studies of how the population changes over time. The  suggestion that X-ray-selected and SZE-selected samples may not have  the same distribution of dynamical states \citep[e.g.][]{pepIX,ros16,and17,lov17} has prompted examination of the relationship between the baryon signatures and the true underlying cluster population. Indeed, the dynamical state may well be as fundamental a characteristic as the mass or the redshift \citep{bar19}. At the same time, \citet{mcd17} found that the new SZE-selected samples suggest that cool cores are in place very early in the history of a cluster, and have not changed in size, density, or total mass up to the present. These authors further suggest that much of what was thought to be cool core growth over time is in fact due to the self-similar evolution of the cluster bulk around this static core.

Here we investigate the properties of the gas density profiles and the core-excised X-ray luminosity, \Lxc, of 31 X-ray-selected clusters at $z<0.2$ and 93 SZE-selected clusters at $z <1.13$. We describe a universal gas density profile for the SZE-selected objects and quantify its variation with redshift and mass. We quantify the radial variation in scaled profiles with respect to the best-fitting evolution and the best-fitting mass dependence. Outside the cores, the median scaled gas density profiles are remarkably similar, showing no dependence on selection.  We obtain the radial variation of the intrinsic scatter in scaled profiles and investigate the evolution of this scatter with redshift. 
We find that \Lxc, measured in the $[0.15-1]\,\Rv$ region\footnote{$R_{\Delta}$ is the radius within which the mean density is $\Delta$ times the critical density at the redshift of the object.}, is an extremely well-behaved mass proxy that does not depend on cluster selection, and shows little evolution beyond self-similar in the broad redshift range of the sample. 

Throughout the paper we assume a flat $\Lambda$CDM cosmology with $\Omega_{\rm m}=0.3$, $\Omega_\Lambda=0.7$, and $H_{0}=70$ km s$^{-1}$ Mpc$^{-1}$. The Sunyaev-Zeldovich flux in units of square arcminutes is denoted $Y_{\rm SZ}$; the quantity $D_{\rm A}^2\, Y_{\rm SZ}$, in units of square megaparsecs, is the spherically integrated Compton parameter within $\Rv$, where $D_{\rm A}$ is the angular diameter distance of the cluster. Unless stated otherwise, logarithmic quantities, including scatter, are given to base $e$, and uncertainties are quoted at the $68\%$ confidence level.  

%
\section{Data and analysis}

\subsection{Dataset}\label{sec:dataset}

The dataset consists of 31 X-ray-selected clusters at $0.05 <z<0.2$ and 93 SZE-selected objects at $0.08<z<1.13$, with six systems in common. The local X-ray-selected dataset is \rexcess\ \citep{boe07,cro08,pra09}. The SZE-selected systems consist of a local sample comprising a subset of 44 objects at $z<0.5$ from the \planck\ Early SZ sample \citep[ESZ;][]{esz}, which \citet{bar19} show is representative of the full ESZ; these were complemented by a further 49 distant objects observed by \xmm\ in a series of three large programmes obtained as part of the M2C project. These cover the redshift ranges $0.5 < z < 0.7$ (LP1, ID 069366, 072378), $0.7 < z < 0.9$ (LP2, ID 078388), and $z>0.9$ (LP3, ID 074440). The LP1 sample was selected from objects detected at a signal-to-noise ${\rm S/N} > 4$ in the \planck\ SZ catalogue, and confirmed by Autumn 2011 to be at $z>0.5$. The LP2 sample consists of clusters with estimated masses $M_{500} > 5\times10^{14}\,\msol$ at $0.7 <z < 0.9$ in the PSZ2 catalogue. The LP3 sample is derived from the five highest-mass objects at $z>0.9$ from the combined  \planck\ and SPT catalogues \citep{bar17, bar18}. Full observation details for the sample can be found in Tables~\ref{tab:obsdetails1},  \ref{tab:obsdetails2},  \ref{tab:obsdetails3}, and  \ref{tab:obsdetails4}.

The left-hand panel of Figure~\ref{fig:sample} shows the distribution of the 118 clusters in the redshift-mass plane. Here and in the following, we group the SZE-selected systems into three sub-samples in the redshift ranges $z < 0.3$ (blue), $0.3 \leq z < 0.6$ (light green), and $z \geq 0.6$ (dark green), containing  $37, 27$, and $29$ objects, respectively. The right-hand panel of Figure~\ref{fig:sample} shows the stacked mass histogram. This plot makes clear that the \rexcess\ sample (light blue) has a lower median mass (${\rm M_{500}} = 2.7\times 10^{14}\, {\rm M}_\odot$) than any of the SZE-selected sub-samples (${\rm M_{500}} = 6.4$, $7.9$, and $5.0 \times 10^{14}\,{\rm M}_\odot$, respectively). The LP2 sub-sample is subject to significant Eddington bias in the \planck\ signal, leading in the most extreme case to an estimated mass of only $6.5 \times 10^{13}\,{\rm M}_\odot$ for PSZ2\,G208.57$-$44.31. We show below that this has a negligible effect on our results.

%
\subsection{Analysis}\label{sec:analysis}

As our aim was to compare the SZE-selected clusters to the low-redshift X-ray-selected \rexcess\ systems, we followed the X-ray data reduction and analysis procedures described in \citet{cro08}, \citet{pra09}, and \citet{pra10}. Event files were reprocessed with the \xmm\ Science analysis System v15 and associated calibration files. Standard filtering for clean events ({\sc Pattern}$<4$ and $<13$ for MOS1/2 and pn detectors, respectively, and {\sc Flag}$=0$) and soft proton flares was applied. The instrumental and particle background was obtained from  custom stacked, recast data files derived from observations obtained with the filter wheel in the CLOSED position (FWC), renormalised using the count rate in a high energy band free of cluster emission.

Vignetting-corrected, background-subtracted [0.3-2]\,keV surface brightness profiles were extracted in annular bins centred on the X-ray peak.  Temperature profiles were produced using the procedures described in \citet{pra10}. These were extracted in logarithmically spaced annular bins centred on the X-ray peak, with a binning of $R_{\rm out}/R_{\rm in} = 1.33=1.5$ depending on data quality. After subtraction of the FWC spectra, all spectra were grouped to a minimum of 25 counts per bin. The FWC-subtracted spectrum of the region external to the cluster was fitted with a model consisting of two {\sc MeKaL} components plus an absorbed power law with a fixed slope of $\Gamma = 1.4$. The spectra were fitted in the $[0.5-10]$ keV range using $\chi^2$ statistics, excluding the $[1.4-1.6]$ keV band (due to the Al line in all three detectors), and, in the pn, the $[7.45-9.0]$ keV band (due to the strong Cu line complex). In these fits the {\sc MeKaL} models were unabsorbed and have solar abundances, and the temperature and normalisations are free parameters; the powerlaw component is absorbed by the Galactic absorption. Since it has a fixed slope, only its normalisation is an additional free parameter in the fit. This best-fitting model was added as an extra component to the annular spectral fits, with its normalisation rescaled to the ratio of the areas of the extraction regions (corrected for bad pixels, chip gaps, etc). In the annular spectral fits, the temperature and metallicity of the cluster component were left free, and the absorption was fixed to the HI value \citep{kal05}. The metallicity was fixed to a value of $Z=0.3\,Z_{\odot}$ when its relative uncertainty exceeded $30\%$.

\subsubsection{Luminosity}

The core-excised X-ray luminosity \Lxc\ was measured in the $[0.15-1]\,\Rv$ region for all objects. Here the only change with respect to the analysis in \citet{pra09} was the use of an updated $M_{500}-Y_{\rm X}$ relation from \citet{arn10} to estimate the relevant masses $M_{500}$ and scaled apertures $R_{500}$. We show below that this change has a negligible impact on the results.  The core-excised X-ray luminosity \Lxc\  was calculated both in the bolometric ($[0.01-100]$\ keV) and soft ($[0.5-2]$\ keV) bands for comparison to previous work. As in \citet{pra09}, the luminosities were calculated from the [0.3-2]~keV band surface brightness profile count rates, using the best-fitting spectral model estimated in the $[0.15-1]\,\Rv$ aperture to convert from count rates to luminosity.  In cases where the surface brightness profile did not extend to $\Rv$ (seven systems), we extrapolated using a power law with a slope measured from the data at large radius. Errors on \Lxc\ take into account the uncertainties in the spectral model, the count rates, and the value of $\Rv$, and were estimated from Monte Carlo realisations in which the luminosity calculation was derived for 100 surface brightness profiles, the profiles and $\Rv$ values each being randomised according to the observed uncertainties. Once obtained, the luminosities were further corrected for point spread function (PSF) effects by calculating the ratio of the observed to PSF-corrected count rates in each aperture (see below).

\subsubsection{Density profiles}

%
\begin{figure*}[!ht]
\begin{center}
\includegraphics[bb=35 560 560 790 ,clip,scale=1.0,angle=0,keepaspectratio,width=1.\textwidth]{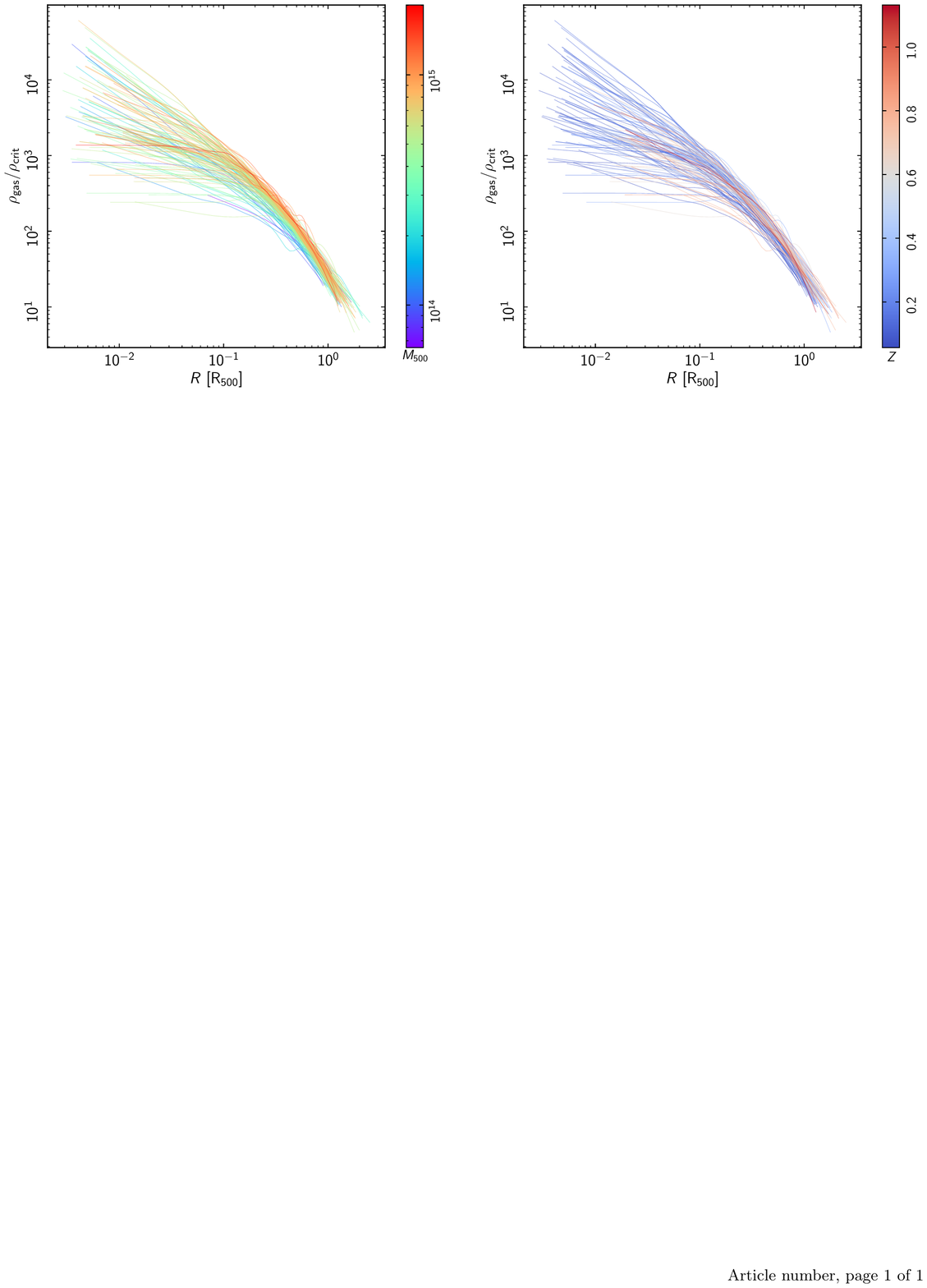}
\end{center}
\caption{\footnotesize Deprojected, PSF-corrected density profiles for 118 galaxy clusters, normalised by the critical density $\rho_{\rm crit}$ and $\Rv$. Fully self-similar clusters would trace the same locus in this plot. The profiles are colour-coded by mass $M_{500}$ in the left-hand panel, and by redshift $z$ in the right-hand panel. There are clear trends with respect to both quantities.
}
\label{fig:xrayprofs}
\end{figure*}
%

The vignetting-corrected background-subtracted [0.3-2]\,keV surface brightness profiles were  used to obtain the deprojected, PSF-corrected density profiles using the regularised, non-parametric technique described in \citet{cro06}, and applied to the \rexcess\ sample in \citet{cro08}. The surface brightness profiles were  converted to gas density by calculating an emissivity profile $\Lambda(\theta)$ in {\sc XSPEC},  taking into account the absorption and instrumental response, and using a parameterised model of the projected temperature and abundance profiles \citep[see e.g.][]{pra03}. The \citet{cro06} method uses the parametric PSF model of \citet{ghi01} as a function of the energy and angular offsets, the parameters of which can be found in EPIC-MCT-TN-011\footnote{\url{http://www.iasf-milano.inaf.it/~simona/pub/EPIC-MCT/EPIC-MCT-TN-012.pdf}} and EPIC-MCT-TN-012\footnote{\url{http://www.iasf-milano.inaf.it/~simona/pub/EPIC-MCT/EPIC-MCT-TN-012.pdf}}. In \citet{bar17}, the deprojected density profiles from \xmm\ observations of a number of clusters obtained using this method were compared to \chandra\ observations, for which the PSF can be neglected. It was shown that the results obtained with the \citet{cro06} method reproduced the deprojected \chandra\ density profiles accurately down to an effective resolution limit of $\sim 5$ arcseconds \citep[Fig. 6 of][]{bar17}.
The gas density $\rhog = n_{\rm e}\times(\mu_{\rm e}\,m_{\rm p})$,  where $n_{\rm e}$ is the electron density measured in X-rays,  $m_{\rm p}$ is the proton mass, and $\mu_{\rm e}=1.148$  is the mean molecular weight  per free electron:
\begin{equation}
\rhog  = 1.92 \times 10^{-24} \left(\frac{n_{\rm e}} {\rm cm^{-3}}\right)~~{\rm g}\ {\rm cm^{-3}}.
\label{eq:rhoconvert}
\end{equation}

%
\section{Gas density profiles}

\subsection{Model}\label{sec:gasmodel}

In the self-similar model, a cluster can be completely defined by only two parameters: its mass, $\MV$, and its redshift, $z$. A fundamental property of this model is the cluster overdensity, $\Delta$, with respect to the reference density of the Universe $\rho_{\rm Uni}(z)$, from which the virial mass $\MV =   \Delta \rho_{\rm Uni}(z) \,(4\pi/3)\, \RV^{3}$, and radius $\RV$, can thereafter be defined. Cluster profiles then exhibit a universal form when the radii are scaled to $\RV$. 

The original, and simplest, self-similar model concerns top-hat spherical collapse in the Standard CDM ($\Omega=1$) cosmology. Here a cluster at redshift $z$ is represented by a spherical perturbation that has just collapsed, with $\rho_{\rm Univ}(z)$ being the critical density of the Universe, $\rho_{\rm c} = 3\,H^2(z) / (8\,\pi\,G)$, and $\Delta=178$. Of course, the hierarchical formation of structure in a $\Lambda$CDM Universe is a very complex dynamical process: objects continuously accrete matter along large-scale filaments,  and there is no strict boundary that would separate a virialised region from the infall zone. The definitions of the mass and the  corresponding overdensity are therefore ambiguous, as is the choice of $\rho_{\rm Univ}(z)$, as one can use either the critical density or the mean density \citep[see][for a review]{vo05}.

Using numerical simulations,  \citet{lau15} showed that the structure of the inner part of clusters that is typically covered by X--ray observations is more self-similar when scaling by fixed overdensities with respect to the critical  density $\rhocz$. The zone in question corresponds to overdensities of $\Delta \gtrsim 200$.  As a scaling  radius, we therefore chose an $\Rv$  corresponding to  $\Delta=500$, the radius within which the mean matter density  is  $\rhov =500\,\rhocz$.  The corresponding total mass within this radius, $\Mv$, is
 \begin{gather}
\Mv  =  (4\pi/3)\,\Rv^3\,\rhov,
\label{eq:m500}
\end{gather}
with 
 \begin{gather}
\rhov    =   500\, \rho_{\rm c}(z) = 4.603 \times 10^{-27}\, E(z)^2\   {\rm g \ cm^{-3}},
\end{gather}
and where $E(z)=[\Omega_{\rm m}(1+z)^3 + \Omega_\Lambda]^{1/2}$ is the evolution of the Hubble parameter with redshift in a flat cosmology. The  scaled gas density profile  expressed  as a function of scaled radius $x$ is then 
\begin{gather}
\rho(x) = \frac{\rhog(R)}{\rhov}  ~~~~{\rm where} ~~~~x  = \frac{R}{\Rv}.
\label{eq:rhosc}
\end{gather}

In the self-similar model, $ \rho(x)$  follows a universal shape and its normalisation is independent of mass and redshift. In such a case  $\langle\rhog\rangle \propto   \rhov $,  where the angle brackets denote the average within $\Rv$, as expected for gas evolution  purely driven by gravitation. Figure~\ref{fig:xrayprofs} shows the scaled density profiles of all 118 systems. If the clusters were perfectly self-similar, they would trace the same locus in this plot. This is clearly not the case. 
The colour-coding by mass and by redshift highlights that at large radius, the scaled profiles of the higher-mass, higher-redshift systems lie systematically above those of lower-mass, lower redshift-objects. These trends suggest a dependence of the scaling on $\Mv$ and/or redshift.  

%
\begin{figure*}[!ht]
\begin{center}
\includegraphics[angle=0,keepaspectratio,width=0.975\textwidth]{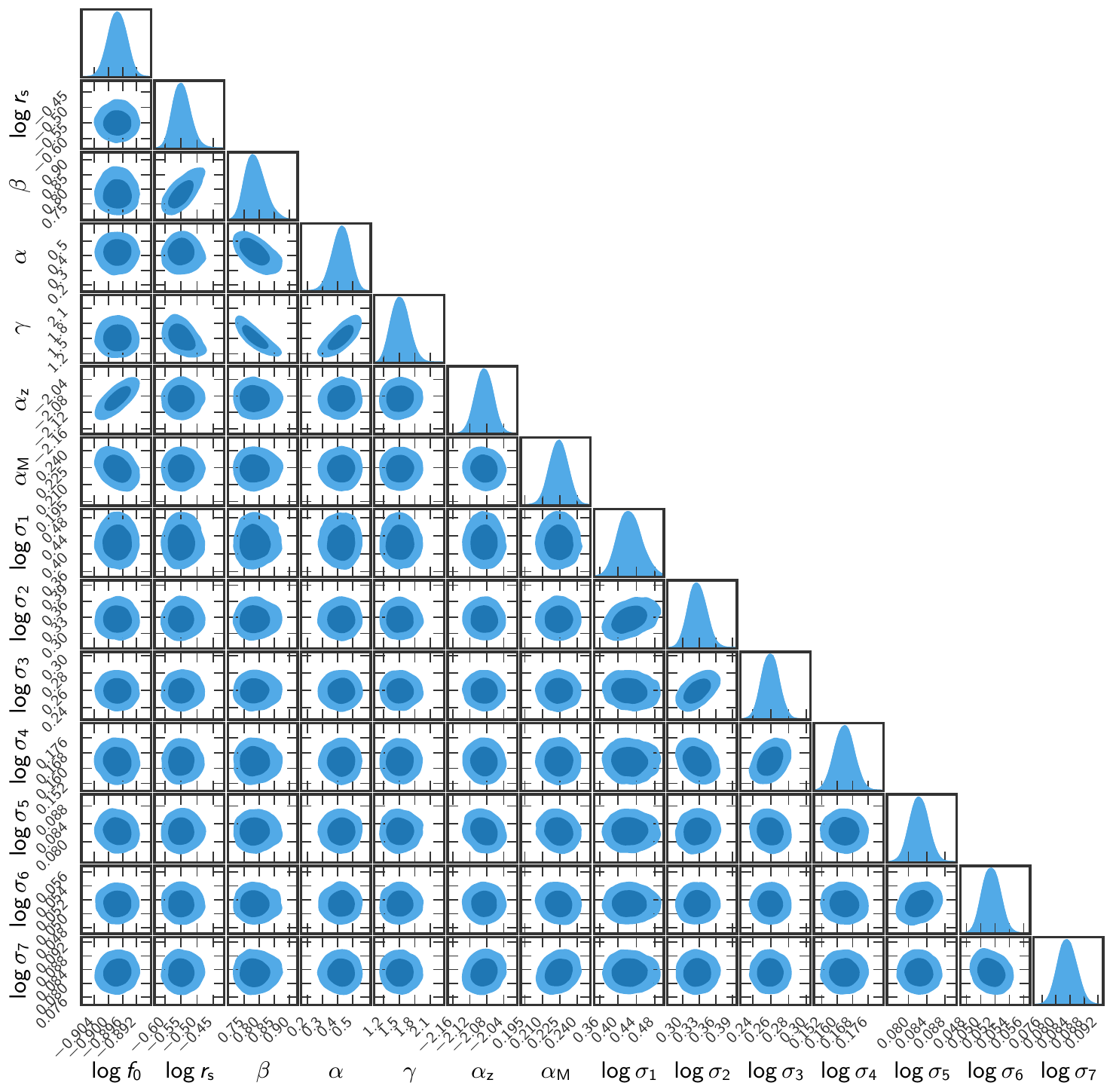}
\end{center}
\caption{\footnotesize Marginalised posterior likelihood for the parameters of the best-fitting density profile model detailed in Sect.~\ref{sec:gasmodel}. 
}
\label{fig:corner}
\end{figure*}
%

%
\begin{figure*}[!ht]
\begin{center}
\includegraphics[bb=35 540 560 790 ,clip,scale=1.0,angle=0,keepaspectratio,width=\textwidth]{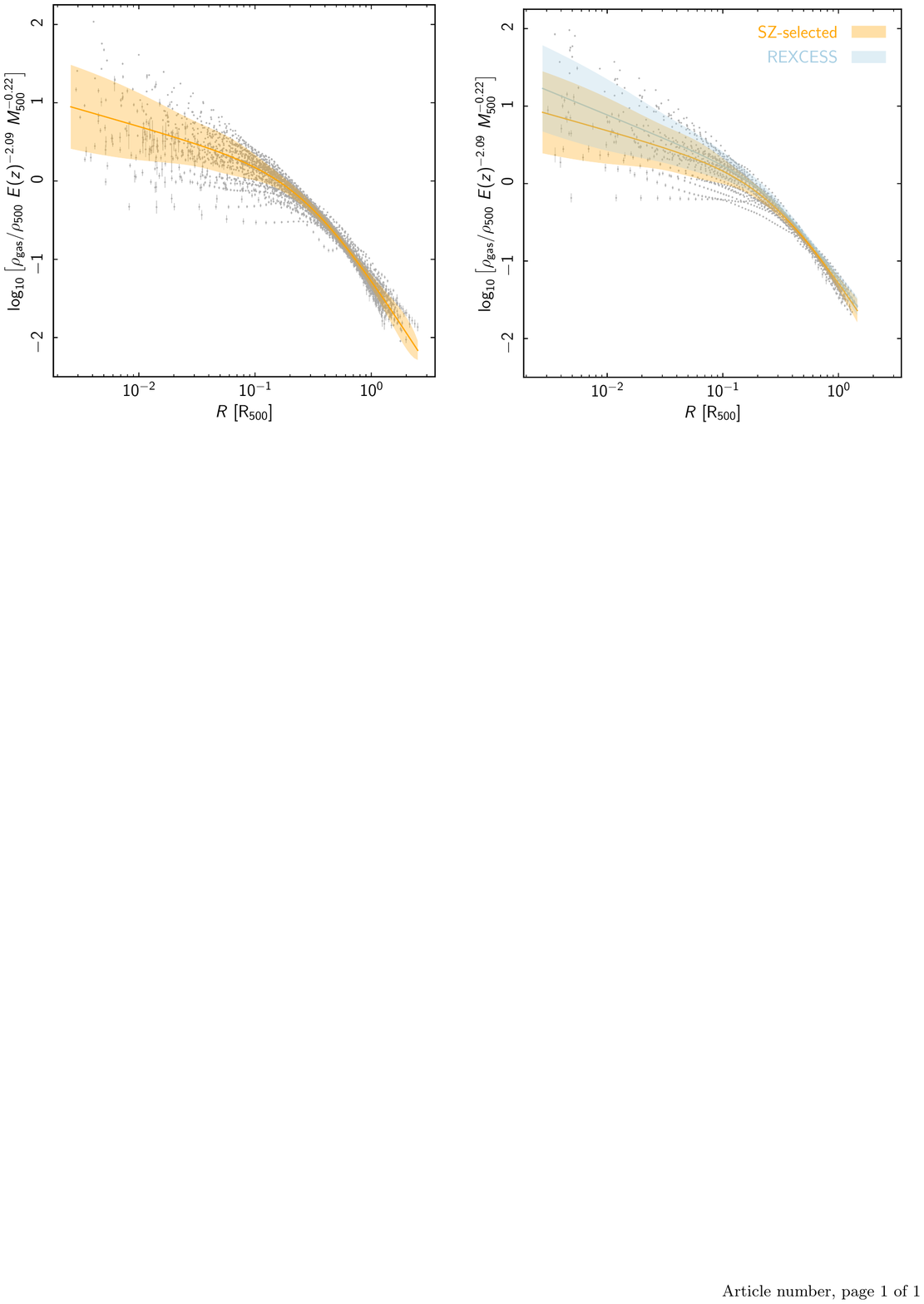}
\end{center}
\caption{\footnotesize The universal cluster ICM density profile. {\it Left:} Scaled density profiles of the SZE-selected clusters (grey points), overplotted with the best-fitting GNFW model with free evolution and mass dependence: $\rho_{\rm gas} / \rho_{500} (R/\Rv) \propto E(z)^{\alpha_{\rm z}}\, M^{\alpha_{\rm M}}$ with $\alpha_{\rm z} = 2.09\pm 0.02$ and $\alpha_{\rm M} =  0.22\pm0.01$  (orange line). The model includes a radially varying intrinsic scatter term (orange envelope). {\it Right:} Comparison of the best-fitting model, defined on the SZE-selected sample, to the best-fitting model for the X-ray-selected \rexcess\ sample (light blue). Here, the points with error bars are the \rexcess\ sample.
}
\label{fig:scmodel}
\end{figure*}

To better understand this dependence, we fitted the observed scaled  profiles with a model consisting of a median analytical profile, the normalisation of which is allowed to vary with $z$ and $\Mv$,  with a radially varying intrinsic scatter. The median profile was expressed as 
\begin{equation}
\rhom(x,z,\Mv)  =  A(z,\Mv) f(x), 
\label{eq:pgnfw}
\end{equation}
where $f(x)$ is the function describing the profile shape. Here we adopted a generalised Navarro-Frenk-White (GNFW) model \citep{nag07}:
\begin{equation}
f(x)   =  \frac{f_0} { (x/\xs)^{\alpha}\left[1+(x/\xs)^\gamma\right]^{(3\beta-\alpha)/\gamma}}, 
\label{eq:gnfw}
\end{equation}
where $\xs$ is the scaling radius, and the parameters $( \alpha,\gamma,3\beta)$ are the central ($x\ll \xs$), intermediate ($x \sim \xs$), and outer ($x \gg x_{\rm s}$) slopes, respectively. The case $\alpha = 0$ and $\gamma=2$ corresponds  to the standard $\beta$ model \citep{cav76}, while the case $\gamma=2$ corresponds to the AB model introduced by \citet{pa02}. The latter was used to model the median density profile of the \rexcess\ sample \citep{Piffaretti11}.

The normalisation is given by the product $f_{0}\,A(z,\Mv)$, where $A(z,\Mv)$ describes the departure from standard self-similarity in terms of a possible mass and/or redshift dependence of the  scaled gas density. For this we assumed a power-law dependence on $\Mv$ and $E(z)$:
\begin{equation}
A(z,\Mv)   =  E(z)^{\az}\, \left[\frac{\Mv}{5 \times10^{14}\,\msol}\right]^{\aM}.  
\label{eq:vary}
\end{equation}
The standard self-similar model corresponds to $(\aM=0, \az=0)$.  We expect $\aM>0$, as it is well established  that the gas mass fraction of local clusters decreases with decreasing mass due to non-gravitational effects \citep[e.g.][]{pra10}.  The model above  allows us to disentangle mass dependence and possible evolution. 

Equations~\ref{eq:gnfw} and~\ref{eq:vary} translate into a gas mass fraction within $\Rv$, $\fgv$,  which varies with mass and redshift as a function of $\aM$ and $\az$:
\begin{eqnarray}\label{eqn:fgasd}
\fgv     =   \frac{\Mgas(<\Rv)}{\Mv} & =&\frac{1}{3} \int_0^1{ \rho(x)  x^2 dx} \\
	& =& f_{0}\, A(z,\Mv)\, I( \xs, \alpha,\gamma,3\beta),
 \end{eqnarray}
where $\Mgas(<\Rv)$ is the gas  mass within $\Rv$ and we have used Eq.~\ref{eq:m500} and Eq.~\ref{eq:rhosc}. The quantity $I( \xs, \alpha,\gamma,3\beta)$  is the three dimensional integral value for $f_{0}=1$, which depends solely on the shape parameters.  

We introduced  a radially varying intrinsic scatter term around the model  profile, assuming a log-normal distribution at each radius. Taking into account measurement errors, the probability of measuring a given scaled gas density $\rho$ at given scaled radius $x$ for a cluster of mass $\Mv$ at redshift $z$ is then
\begin{gather}
p\,(\rho \vert x, z, \Mv)  =  \mathcal{N}\,[ \log{\rhom\,(x,z,\Mv)}, \sigma^2(x)] \\  
\sigma^2(x)  = \sigma^2_{\rm int}(x)    +  \sigma^2_{\rm stat},
\end{gather}
where $\mathcal{N}$ is the log-normal distribution.The variance term, $\sigma^2(x)$, is  the quadratic sum of the statistical error,  $\sigma_{\rm stat}$ on the measured $\log{\rho}$ and of the intrinsic scatter on $\log{\rhom}$ at radius $x$,  $\sigma_{\rm int}(x)$. 

We expect the intrinsic scatter  to increase towards the centre, as observed in Fig.~\ref{fig:xrayprofs}, due to the increasing effect of non-gravitational physics on the density profiles.  \citet{ghi19} studied the intrinsic scatter of  massive local clusters in the XCOP sample, modelling the radially varying scatter $\sigma_{\rm int}(x)$ with a log-parabola function. However, we found that such an analytical form significantly overestimates the scatter in the inner core, which was not covered by their data. To allow for more freedom we used a non-analytical form for the intrinsic scatter, where $\sigma_{\rm int}(x)$ is defined at $n$ equally spaced points in $\log(x)$ in the typical observed radial range, $[x_{\rm {min}}$--$x_{\rm_{max}}]$. The scatter, $\sigma_{\rm int}(x)$ at other radii is computed by spline interpolation.  We used $n=7$ between $x_{\rm min}=0.01$ and $x_{\rm max}=1$. 

The likelihood of  a  set of scaled density profiles measured  for  a sample of  $i=1, N_{\rm c}$ clusters of mass $\Mvi$ and redshift $\zi$ is:
\begin{gather}
\mathcal{L}  = \prod_{i=0}^{N_{\rm c}} \prod_{j=0}^{ N_{\rm R}[i]} p\,(\rhoij \vert \xij, \zi,\Mvi), %
\end{gather}
where   $N_{\rm R}[i]$ is the number of  points  of the  profile of cluster $i$, and the quantity $\rhoij= \rhog[i,j]/\rhov(\zi)$ is the scaled density  measured at each scaled radius $\xij=r[i,j]/\Rv(\zi,\Mvi)$, with $\rhog[i,j]$  and $r[i,j]$ being  the physical gas density and radius.  The statistical error on $\log{\rhoij}$ is $\sigma_{\rm stat,i,j}$. 

We  fitted the data (i.e. the set of observed $\rhoij$) using Bayesian maximum likelihood estimation with Markov Chain Monte Carlo (MCMC) sampling.  Using the \texttt{emcee} package developed by  \citet{for13}, we maximised the log of the likelihood, which reads (up to an additive constant)
\begin{gather}
\ln{\mathcal{L}}  = -0.5 \sum_{i,j} \left[ \ln{\sigdij }  + \frac{ \left(\log{\rhoij}- \log{\rhomij} \right)^2  } { \sigdij} \right] \\ 
\rhomij  =  \rhom(\xij,\zi,\Mvi) \\
  \sigdij  =  \sigma^2_{\rm int}(\xij)    +  \sigma^2_{\rm stat,i,j}.
\end{gather}
The fit marginalises over a total of fourteen parameters: four describing the shape of the median profile $(\xs, \alpha,\gamma,3\beta)$, a global normalisation, $f_{0}$, the slopes $\aM$ and $\az$ that describe the non-standard mass and evolution dependences, and seven additional parameters describing the intrinsic scatter profile. We used flat priors on all parameters.

%

\subsection{Results}\label{sec:modres}

To establish a baseline, we fitted the model described above to the 93~SZE-selected systems. The resulting best-fitting model is
\begin{gather}
\rhom(x, z,\Mv)  =  A(z,\Mv) f(x), 
\label{eqn:fullmod}
\end{gather}
with
\begin{gather}\label{eqn:bestmod1}
A(z,\Mv) = E(z)^{2.09\pm0.02} \times \left[\frac{\Mv}{5 \times 10^{14}\, h_{70}^{-1}\, {\rm M}_{\odot}}\right]^{0.22\pm0.01}
\end{gather}
and
\begin{gather}
f(x)   =  \frac{f_0} { (x/\xs)^{\alpha}\left[1+(x/\xs)^\gamma\right]^{(3\beta-\alpha)/\gamma} },
\label{eqn:bestmod2}
\end{gather}
where 
\begin{gather*}
f_0 = 1.20 \pm 0.15, \\
\xs = 0.28\pm0.01, \\
\alpha = 0.42\pm0.06, \\ 
\beta = 0.78\pm0.03,\ {\rm and}\\
\gamma = 1.52\pm0.16. 
\end{gather*}
Figure~\ref{fig:corner} shows the marginalised posterior likelihood for the parameters of the best-fitting density profile model detailed in Sect.~\ref{sec:gasmodel}. All parameters are well constrained: in particular, we note that the mass and evolution parameters $\alpha_{\rm M}$ and $\alpha_{\rm z}$ do not show any degeneracies, implying that we clearly separate the mass and redshift effects.

The left-hand panel of Fig.~\ref{fig:scmodel} shows the density profiles of the SZE-selected clusters together with the best-fitting model. The intrinsic scatter term is represented by the orange envelope; the numerical values for this term are given in Table~\ref{tab:scatter}. The right-hand panel of Fig.~\ref{fig:scmodel} shows the best-fitting model for the SZE-selected systems compared to the profiles from the X-ray-selected \rexcess\ sample. The agreement is excellent beyond the core; in the inner regions, there is a hint that the X-ray-selected systems may show more dispersion. We will return to this point below in Sect.~\ref{sec:centreg}.

\begin{table}[]
\caption[]{\footnotesize Numerical values for the best-fitting intrinsic scatter term, measured at seven equally spaced points in $\log(R/\Rv)$, in the range $[0.01-1]\,\Rv$. }
\label{tab:scatter}
\centering

\begin{tabular}{@{}ll@{}}

\toprule
\toprule
Radius &  $\sigma_{\rm int}$ \\

\midrule

0.010 & $0.98\pm0.06$ \\
0.021 & $0.75\pm0.03$ \\
0.046 & $0.60\pm0.02$ \\ 
0.100 & $0.38\pm0.01$ \\
0.215 & $0.19\pm0.01$ \\
0.464 & $0.12\pm0.01$\\
1.000 & $0.19\pm0.01$\\
\bottomrule
\end{tabular}
\end{table}


%
\section{Luminosity scaling relations}

We now turn to the scaling relation between \Lxc\ and the mass $\Mv$. The bolometric X-ray luminosity of a cluster can be written \citep{ae99}
\begin{gather}
L(T) = f_{\rm gas}^2(T) [M(T) \Lambda(T)] \hat{Q}(T),
\label{eqn:lxeqn}
\end{gather}
where $f_{\rm gas} = M_{\rm gas} / M$ is the gas mass fraction, and $\Lambda(T)$ is the cooling function. The quantity $\hat{Q}(T)= \langle \rho_{\rm gas}^2 \rangle / \langle \rho_{\rm gas}\rangle^2$ is a dimensionless structure factor that depends on the
spatial distribution of the gas density (e.g. clumpiness at small scale, shape at large scale, etc.). Further assuming (i) virial equilibrium of the gas in the dark matter potential  [$M \propto T^{3/2}$]; (ii) simple Bremsstrahlung emission [$\Lambda(T) \propto T^{1/2}$]; (iii) similar internal structure [$\hat{Q}(T) = {\rm const.}$]; (iv) a constant gas mass fraction [$f_{\rm gas}^2(T) = {\rm const.}$], the standard self-similar relation between bolometric X-ray luminosity and mass, $L_{\rm X} \propto M^{4/3}$, can be obtained. Similar arguments can be used to obtain the soft-X-ray luminosity-mass relation of $L_{\rm X} \propto M$.

%
\begin{figure}[!t]
\begin{center}
\includegraphics[angle=0,keepaspectratio,width=0.975\columnwidth]{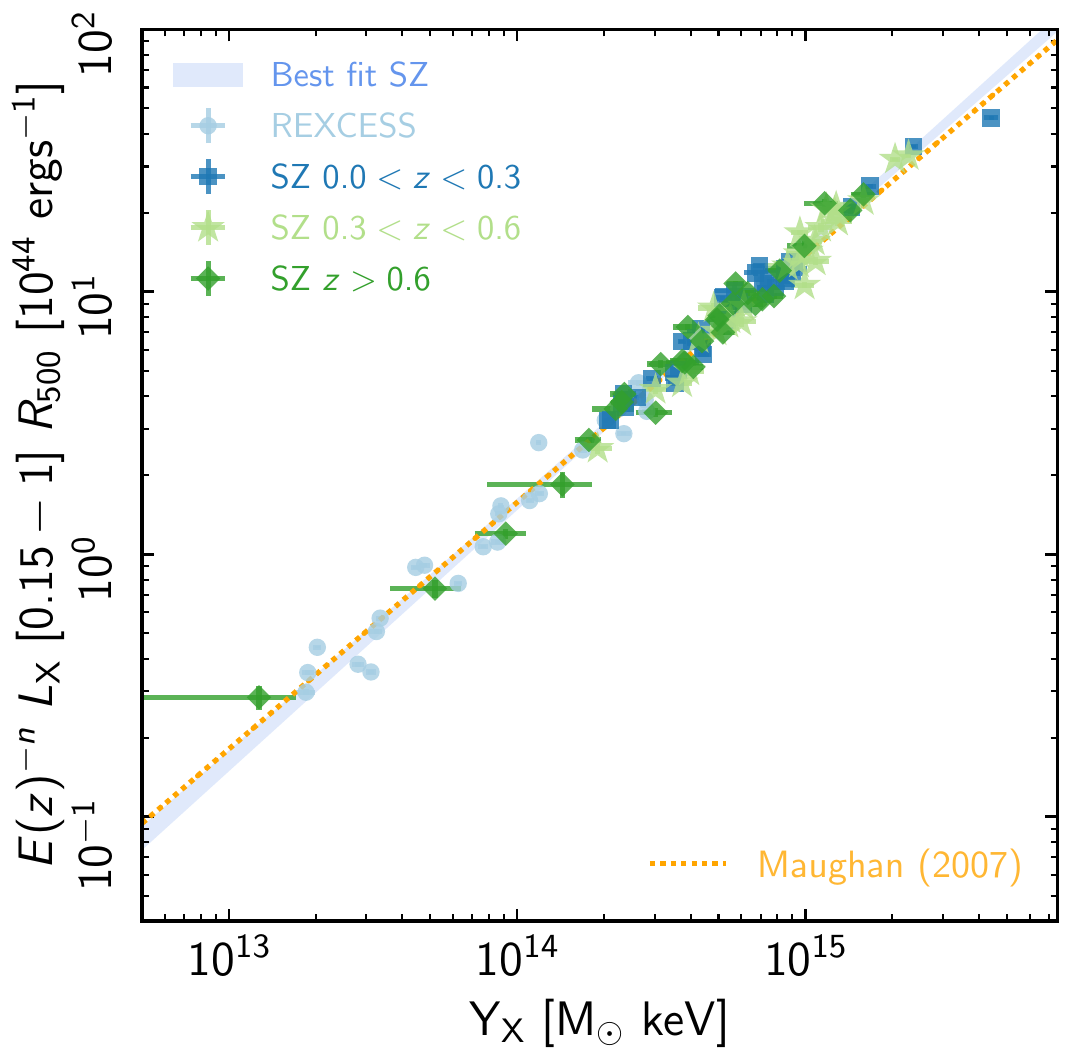}
\end{center}
\caption{\footnotesize Relation between \Lxc\ and $Y_{\rm X}$. The blue envelope is the best-fitting relation given in Eqn.~\ref{eqn:lxyx}, and the results from \citet{mau07} are also shown for comparison.}
\label{fig:LxYx}
\end{figure}

%
\begin{figure*}[]
\begin{center}
\includegraphics[bb=35 540 560 785 ,clip,scale=1.,angle=0,keepaspectratio,width=0.975\textwidth]{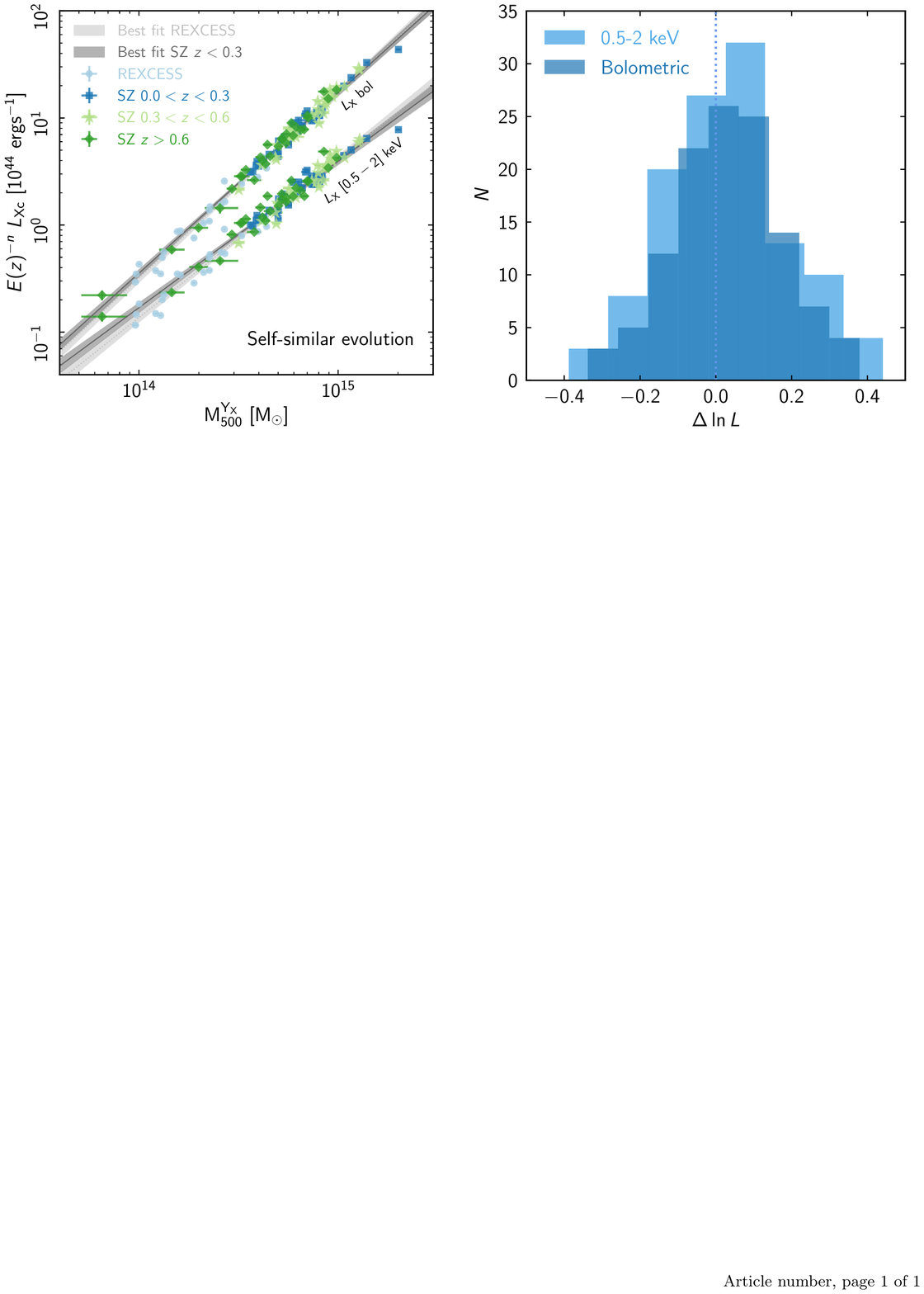}
\end{center}
\caption{\footnotesize 
Relation between the core-excised X-ray luminosity \Lxc\ and mass estimated from the $Y_{\rm X}$ proxy, for the bolometric and soft-band luminosities of 118 systems. {\it Left:} Data points with the best fitting relation to the X-ray-selected \rexcess\ sample with the evolution factor fixed to the self-similar value of $n=-2$ (grey envelope). The dark grey envelope shows the best fitting relation to the 37 SZE-selected systems at $z<0.3$ with $n=-2$. {\it Right:} Histogram of the log space residuals from the best fitting relation to the SZE-selected objects at $z<0.3$.
}
\label{fig:lmrelsss}
\end{figure*}

%
\begin{figure*}[]
\begin{center}
\includegraphics[bb=35 540 560 785 ,clip,scale=1.,angle=0,keepaspectratio,width=0.975\textwidth]{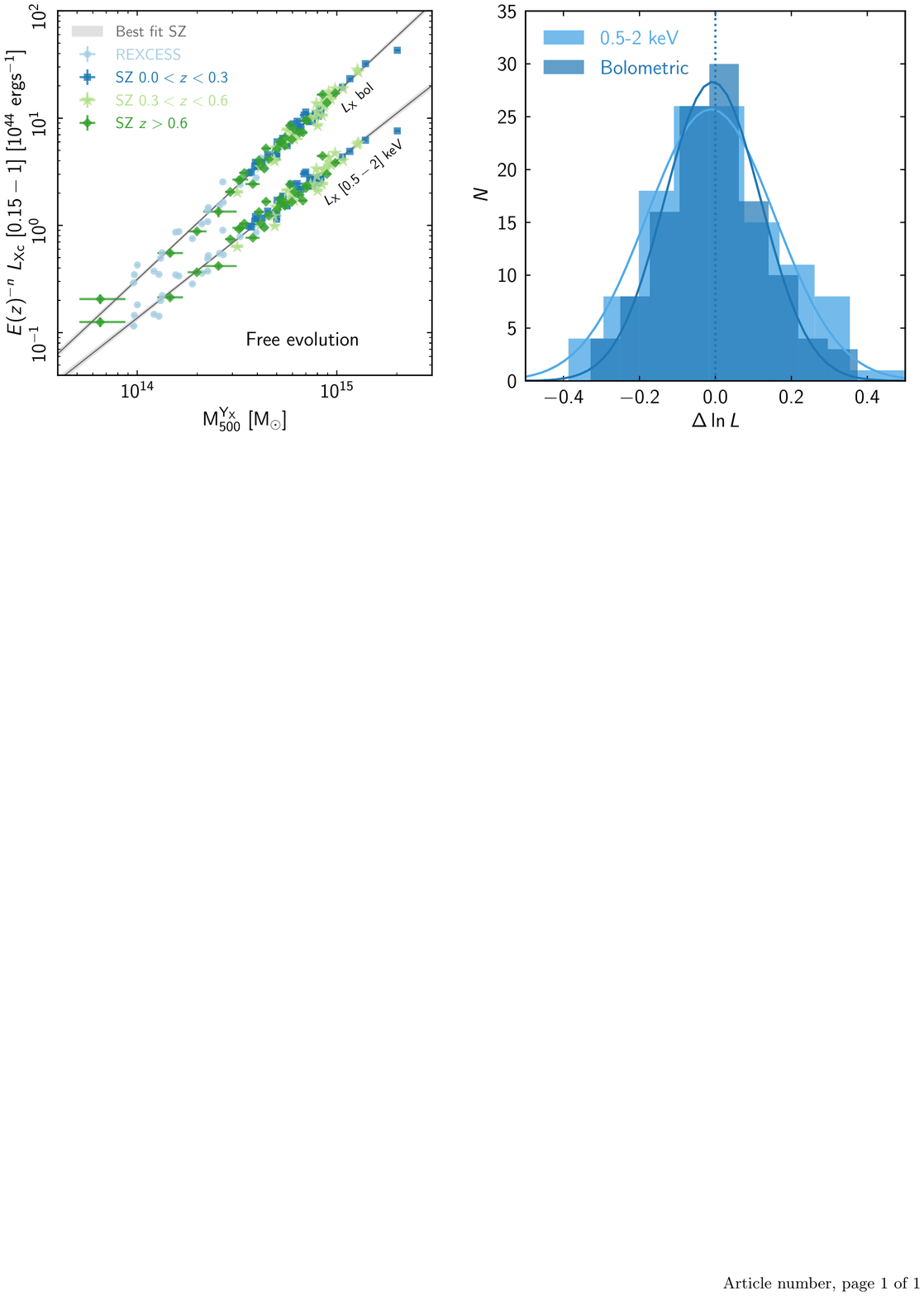}
\end{center}
\caption{\footnotesize 
Relation between the core-excised X-ray luminosity \Lxc\ and mass estimated from the $Y_{\rm X}$ proxy, for the bolometric and soft-band luminosities of 118 systems. {\it Left:} Best fitting relation (grey envelope) to the full sample (data points) with the evolution factor $n$ left free to vary. The best-fitting  values of $n$ are given in Table~\ref{tab:lmrels}.  {\it Right:} Histogram of the log space residuals from the best fitting relation. Solid lines show the best-fitting Gaussian distributions with $\sigma$ corresponding to the best-fitting intrinsic scatter in log space (Table~\ref{tab:lmrels}).
}
\label{fig:lmrelsnss}
\end{figure*}
%
\begin{table*}[]
\caption[]{\footnotesize Fits to the core-excised X-ray luminosity \Lxc$ - M$ relation. \Lxc\ is measured in the $[0.15-1]\,\Rv$ region. }
\label{tab:fits}
\centering

\begin{tabular}{@{}llcccccc@{}}

\toprule
\toprule
Relation &  Selection & Redshift & $n$ & $L_0$ & $A_{\rm L}$                                     & $B_{\rm L}$                       & $\sigma_{\rm \ln{L}}$ \\
              &                  &                &       & $(10^{44}$ erg s$^{-1})$  &     &    &                                  \\
\midrule
bolometric & \rexcess &                $0.05<z <0.20$ &   $-7/3$ & $5$ & $0.70\pm0.03$ & $1.73\pm0.05$ & $0.16\pm0.03$  \\
                       & SZ         & $z<0.3$ &  $-7/3$ & $5$ & $0.73\pm0.02$ & $1.69\pm0.05$ & $0.09\pm0.02$ \\
                          & all  SZ  & $z> 0.05$ &     $-2.50\pm0.09$ & $5$ & $0.70\pm0.02$ & $1.74\pm0.02$ & $0.12\pm0.01$  \\
\\
0.5-2\, keV & \rexcess	& $0.05<z<0.20$        & $-2$    & $1$ & $1.06\pm0.04$ & $1.48\pm0.05$ & $0.17\pm0.03$ \\
    		& SZ 		&  $z<0.30$  		   & $-2$   & $1$ & $1.12\pm0.04$ & $1.37\pm0.06$ & $0.13\pm0.02$  \\
		& all SZ        & $z>0.05 $       & $-2.23\pm0.10$   &  $1$ & $1.05\pm0.03$ & $1.46\pm0.03$ & $0.15\pm0.02$
                          \\
\bottomrule
\end{tabular}
\tablefoot{The data were fitted in log-log space with a relation of the form $E(z)^n\,(L/L_0)  = A_{\rm L}\, (M/4\times10^{14}\,{\rm M}_{\odot})^{B_{\rm L}}$. \label{tab:lmrels} }
\end{table*}

\subsection{Fitting method}

We fitted the data with a power-law relation of the form
\begin{equation}
E(z)^n\, L/L_0= A_{\rm L}\,(X/X_0)^{B_{\rm L}},\label{eqn:lm}
\end{equation}
where $L_0 = 1\times 10^{44}$ erg~s$^{-1}$ and $5\times 10^{44}$ erg~s$^{-1}$ for the soft and bolometric bands, respectively, and $X_0=5\times10^{14}\, {\rm M}_{\odot}\,{\rm keV}$ and $4\times 10^{14}\, {\rm M}_{\odot}$ for $Y_{\rm X}$ and $M$, respectively.  Fitting was undertaken using linear regression in the log-log plane, taking uncertainties in both variables into account, and including the intrinsic scatter. We fitted the data using a Bayesian maximum likelihood estimation approach with Markov Chain Monte Carlo (MCMC) sampling. We write the likelihood as defined by  \citet{rob15}
\begin{gather}    
     \ln{\mathcal {L}}  =   \sum_{i=1}^N \left[    \ln{  \frac{ B_{\rm L}^2 + 1}{\sigma_i^2} }  - \frac{\left(  \ln{(L_i/L_0)}- \ln{A_{\rm L}} -B_{\rm L} \ln{(X_i/X_0)}   \right)^2  } {\sigma_i^2} \right],
\end{gather}
with
\begin{gather}
 \sigma_i^2  =  \sigma^2    + B_{\rm L}^2\sigma_{X,i}^2  + \sigma_{L,i} ^2 
\end{gather}
and the intrinsic scatter, $ \sigma^2$, as a free parameter. MCMC sampling was undertaken using the \texttt{emcee} package developed by \citet{for13}, with flat priors in the ranges $[-2.0,3.0]$ and  
$[0.0,1.0]$ for $B_{\rm L}$ and $\sigma$, respectively. The results, reported in Table~\ref{tab:lmrels}, were compared to those obtained with the \verb?LINMIX? \citep{kel07} Bayesian regression package:  these were indistinguishable and so are not reported here.


%
\begin{figure*}[!ht]
\begin{center}
\includegraphics[bb=35 540 560 785 ,clip,scale=1.,angle=0,keepaspectratio,width=0.975\textwidth]{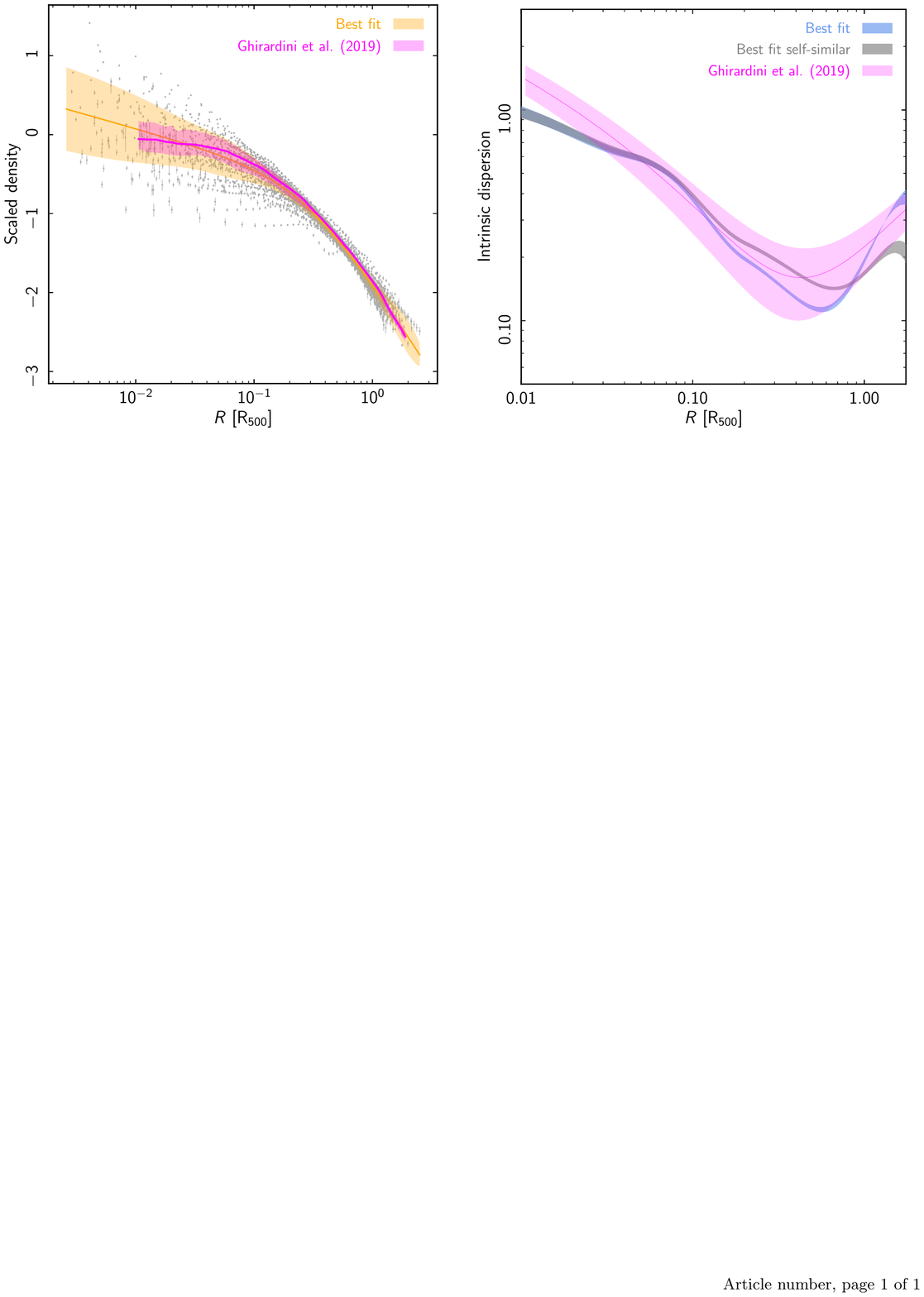}
\end{center}
\caption{\footnotesize Scaled density profiles. {\it Left:} Scaled density profiles (points) and best-fitting model (orange envelope) for SZE-selected systems in our sample compared to the median and $68\%$ dispersion from the X-COP sample \citep[][magenta envelope]{ghi19}. {\it Right:} Comparison of best fitting intrinsic scatter model (blue) with that found by \citet[][magenta]{ghi19}. The best-fitting intrinsic scatter obtained from our sample when the evolution factor is forced to the self-similar value of $E(z)^2$ is also shown in grey.  
}
\label{fig:compghi}
\end{figure*}

\subsection{Results}
\label{sec:lxmres}

\subsubsection{\Lxc$- Y_{\rm X}$}

We first fitted the relation between the bolometric \Lxc\ and the mass proxy $Y_{\rm X}$ for the SZE-selected systems only. With the evolution term left free, the best-fitting relation, shown in Fig.~\ref{fig:LxYx}, is
\begin{gather}
E(z)^{-1.79\pm0.08}\,(L_{\rm Xc}/L_0) = (1.49\pm0.03)\,(Y_{\rm X}/Y_0)^{0.97\pm0.02}\label{eqn:lxyx},
\end{gather}
with an intrinsic scatter of $\sigma_{L_{\rm Xc} | Y_{\rm X}} = 0.09\pm0.01$. This result yields evolution and mass dependences that are in excellent agreement with the self-similar predictions of $-9/5$ and 1.0, respectively. It is also in excellent agreement with the \rexcess\ only results of \citet{pra09} and that of \citet{mau07}, the latter of which was estimated from {\it Chandra} data and is overplotted on the figure.


%
\subsubsection{\Lxc$- M$: Low redshift with fixed evolution}

We initially fixed the evolution factor, $n$, to the self-similar values of $-2$ and $-7/3$ for the soft and bolometric bands, respectively. 
We first fitted the bolometric \Lxc--$M$ for the \rexcess\ data only, using a mass pivot of $2\times 10^{14}\,\msol$, as used by \citet{pra09}. The resulting normalisation, $A_{\rm L} = 1.06\pm0.04$, slope $B_{\rm L} = 1.73 \pm 0.05$, and intrinsic scatter $\sigma_{\ln{L_{\rm Xc}}} = 0.16\pm 0.03$, are in excellent agreement with those found by \citet{pra09} using orthogonal Bivariate Correlated Errors and intrinsic Scatter \citep[BCES;][]{akr96} fitting. Similarly good agreement was found for the soft-band  \Lxc--$M$ relation, showing that the scaling relation parameters that are obtained for the X-ray-selected are robust to the change in underlying $M-Y_{\rm X}$ relation used to estimate the mass, and also to differences in fitting method. 

We then fitted the SZE-selected clusters at $z< 0.3$ (37 systems). The results are given in Table~\ref{tab:lmrels} and show that the normalisation and slope for this sub-sample are in agreement within $1\sigma$ with those found for \rexcess. This indicates that there is no difference in the scaling relation between the local X-ray and SZE-selected samples, once the core region has been excised. There is a slight hint that the intrinsic scatter of the SZE-selected sample about the best-fitting relation is lower than that for \rexcess, although this is only a $\sim 2\sigma$ effect for the bolometric luminosity, and is less significant for the soft-band.  The data and best-fitting relations, including the $1\sigma$ scatter envelopes, are shown in the left-hand panel of Fig.~\ref{fig:lmrelsss}.


\subsubsection{\Lxc$- M$: Free evolution}

The right-hand panel of Fig.~\ref{fig:lmrelsss} shows the histogram of the residuals of the full SZE-selected sample  (93 systems, $0.08 < z < 1.13$) with respect to the best-fitting relation to the systems at $z<0.3$. The peak is offset by $\Delta \ln{L} \lesssim 0.1$, indicating that some evolution beyond self-similar is in fact needed.

We then fitted the full SZE-selected sample with a power-law relation, including a free evolution factor, $n$. The results are given in Table~\ref{tab:lmrels} and the best-fitting relations are shown in the left-hand panel of  Fig.~\ref{fig:lmrelsnss}; the right-hand panel of Fig.~\ref{fig:lmrelsnss} shows the residual histograms. The latter are well-centred on zero. The best-fitting evolution terms,  $n=-2.23\pm0.09$ for the soft-band and  $n=-2.50\pm0.10$ for the bolometric luminosity, suggest that stronger than self-similar evolution is significant at the $\sim 2 \sigma$ level.

\begin{figure*}[!ht]
\begin{center}
\includegraphics[bb=35 620 560 785 ,clip,scale=1.0,angle=0,keepaspectratio,width=\textwidth]{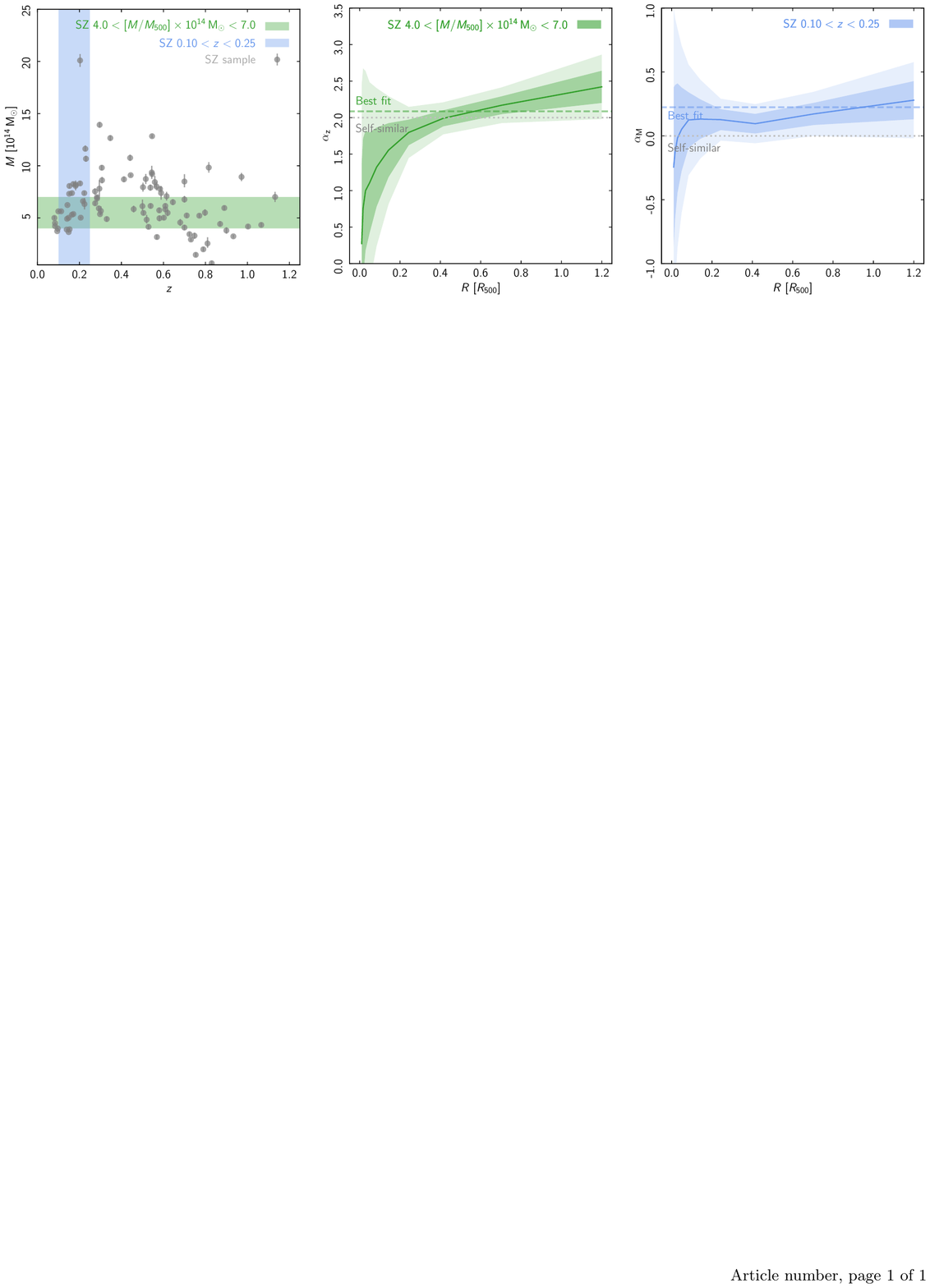}
\end{center}
\caption{\footnotesize Deviations from the average scaling with radius. {\it Left:} Redshift-mass distribution of the SZE-selected sample used in this work. The shaded regions indicate cuts for two sub-samples: a large redshift range at nearly fixed mass, and a large mass range at nearly fixed redshift. 
{\it Middle:} Degree to which the radial ICM density profile evolves as a function of redshift at nearly fixed mass. The dotted line shows the self-similar expectation ($\alpha_{\rm z} = 2$). The dashed line shows the best-fitting evolution, which varies from slower than self-similar in the centre ($\alpha_{\rm z} \sim 0.3$) to faster than self-similar around $\Rv$ ($\alpha_{\rm z} \sim 2.4$). Envelopes show the $1$ and $2\sigma$ uncertainties. 
{\it Right:} Degree to which the radial ICM density profile scales with mass at nearly fixed redshift. The dotted line shows the self-similar expectation ($\alpha_{\rm z} = 0$). The dashed line shows the best-fitting mass dependence of $\alpha_{\rm z}=0.22$. The scaled density at nearly fixed redshift does not depend on radius.
}
\label{fig:evolwithmassandz}
\end{figure*}


\section{Discussion}

\begin{figure*}[!ht]
\begin{center}
\includegraphics[bb=35 540 560 785 ,clip,scale=1.0,angle=0,keepaspectratio,width=0.97\textwidth]{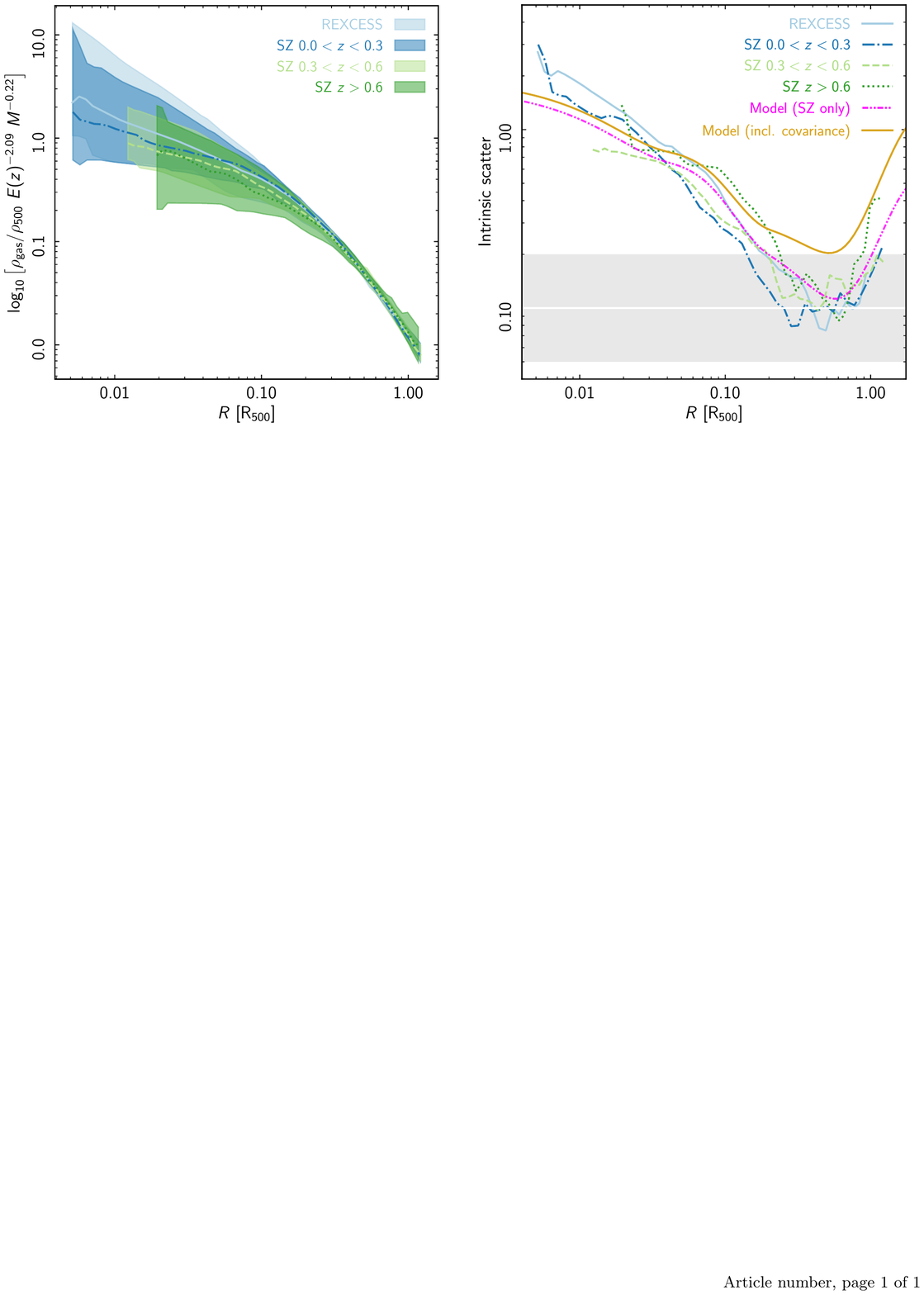}
\end{center}
\caption{\footnotesize Scaled profiles and scatter. {\it Left:} Median scaled profiles (solid lines) and $68\%$ scatter (envelopes) for \rexcess\ and the SZE-selected sample split into three redshift bins. Beyond $\sim 0.2\,\Rv$ the scaled profiles are almost indistinguishable.  {\it Right:} Radial profile of the intrinsic scatter for the various sub-samples. The best-fitting intrinsic scatter model obtained from the SZE-selected sample is also shown. Intrinsic scatter is less than $20\%$ between $0.2 \lesssim \Rv \lesssim 1.0$. The gold line shows the model intrinsic scatter profile corrected for the covariance between $M_{\rm gas}$ and $\Rv$ (see Sect.~\ref{sec:dispcov}), which results in a suppression of the scatter by a factor of about two at $\Rv$.  
}
\label{fig:envscatter}
\end{figure*}

\subsection{Gas density profiles}
\subsubsection{Comparison with previous work}

%
%
Pioneering work on parametric models of scaled density profiles \citep{neu99} obtained from ROSAT allowed the dispersion in radial slopes to be constrained. \citet{cro08} studied the scaled density profiles of the \rexcess\ sample, obtaining for the first time constraints on the radial dependence of the intrinsic scatter. The scaled density profile of the X-COP sample, obtained assuming a self-similar evolution factor  $E(z)^2$, was presented in \citet{ghi19}. Figure~\ref{fig:compghi} compares the scaled density profiles and the best-fitting model from our SZE-selected sample to their median scaled density profile and $68\%$ dispersion. The agreement is good out to the maximum X-COP radius of $\sim2\,\Rv$, although with subtle differences in the inner regions ($R<0.1\,\Rv$). Their profile is less peaked, likely due to their not having corrected the density profiles for PSF effects, and has a smaller dispersion than our sample, which may be linked to their more limited mass coverage. The right-hand panel compares the intrinsic scatter measurements, which also agree quite well, although the scatter of the present sample is better constrained. The best-fitting intrinsic scatter obtained from our sample when the evolution factor is forced to the self-similar value of $E(z)^2$ is also shown in grey. 

We can also compare to the results obtained by \citet{man16}, who modelled the evolution and mass dependence of the scaled density profiles of a morphologically relaxed cluster sample of 40 systems at $0.08 <z <1.06$. While their evolution dependence of $2.0\pm0.2$ is in agreement with our results, they find a mass dependence that is consistent with zero ($0.03\pm0.06$). The difference with respect to our results may be due simply to cluster selection. They studied dynamically relaxed, hot systems, for which the mass leverage is more limited. Once scaled, they found a scatter in scaled density of $\lesssim 20\%$ at $R_{2500}$. For the typical mass of the present sample, $R_{2500} \approx 0.45\,\Rv$, where our intrinsic scatter measurements are in good agreement with theirs. 

\subsubsection{Radial dependence of the scaling}

Once the best fitting model was obtained, we quantified how well the model represents the data by calculating the variation of scaled density at different scaled radii. To better disentangle redshift and mass evolution, we extracted two sub-samples: one covering a large redshift at nearly constant mass, and another covering a large mass range at nearly constant redshift. These sub-samples are illustrated in the $z-M$ plane by the orange and blue regions in the left-hand panel of  Fig.~\ref{fig:evolwithmassandz}. We defined ten radial bins in terms of $x=r/\Rv$ and measured the gas density in each bin. We then scaled the density by the best-fitting model and fitted a function of the form $\rho_{\rm gas} (x) / \rho_{\rm crit,0}  \propto E(z)^{\alpha_{\rm z}}\, M^{\alpha_{\rm M}}$. For a self-similarly evolving population, the gas density scales with the critical density $\rho_{\rm crit}$ and there is no mass dependence, and so $\alpha_{\rm z}=2$ and $\alpha_{\rm M}=0$.

The middle- and right-hand panels of  Fig.~\ref{fig:evolwithmassandz} show the degree to which the two sub-samples vary from self-similar scaling and from the best-fitting model scaling, as a function of scaled radius. Uncertainties are large because of the reduced number of data points and the scatter in the data. At nearly constant mass the overall variation with redshift is slightly greater than self-similar. However, the density evolves differently in the core and in the outer regions. The density evolution is consistent with zero in the core: $\alpha_{\rm z} = 0.28\pm 1.10$ at $x = 0.01$,  a value that is in good agreement  the result found by \citet{mcd17} although with large uncertainties. However, the density in the outer regions appears to evolve more strongly than self-similar:  $\alpha_{\rm z} = 2.42\pm 0.22$ at $x = 1.20$, a result that is significant at slightly more than $1\sigma$.
At nearly fixed redshift, the density varies with mass in agreement with the $\alpha_{\rm M} = 0.22\pm0.01$ scaling established above. At $\Rv$, this mass dependence is significant at $\sim 2\sigma$, but does not depend on radius.


\subsubsection{Median and scatter of scaled profiles}

We now turn to the ensemble properties of the scaled density profiles. The left-hand panel of Figure~\ref{fig:envscatter} shows the median scaled profile, obtained in the log-log plane, and $68\%$ dispersion for \rexcess\ X-ray-selected sample and the SZE-selected sample split into three redshift bins. It is clear that once scaled, the four sub-samples are remarkably similar beyond $0.2\,\Rv$. In the core region, the median central density decreases progressively with redshift. The median scaled central densities of the \rexcess\ and of the SZE-selected sample at $z<0.3$ are virtually indistinguishable. We will return to the central regions in Sect.~\ref{sec:centreg} below.

The radial variation of the intrinsic scatter about these median profiles is quantified in the right-hand panel of Figure~\ref{fig:envscatter}, together with that of the best-fitting intrinsic scatter model obtained above (from the SZE-selected clusters only). The intrinsic scatter of this model falls below $20\%$ in the radial range $0.2 \lesssim \Rv \lesssim 1.0$. There is excellent agreement between the best-fitting intrinsic scatter model and the observed profiles, which all follow broadly the same trend with scaled radius: a steep decrease with a minimum at $\sim 0.5-0.7\Rv$, followed by an increase towards larger radii.  
%
\begin{figure*}[!ht]
\begin{center}
\includegraphics[bb=35 620 560 785
,clip,scale=1.,angle=0,keepaspectratio,width=0.975\textwidth]{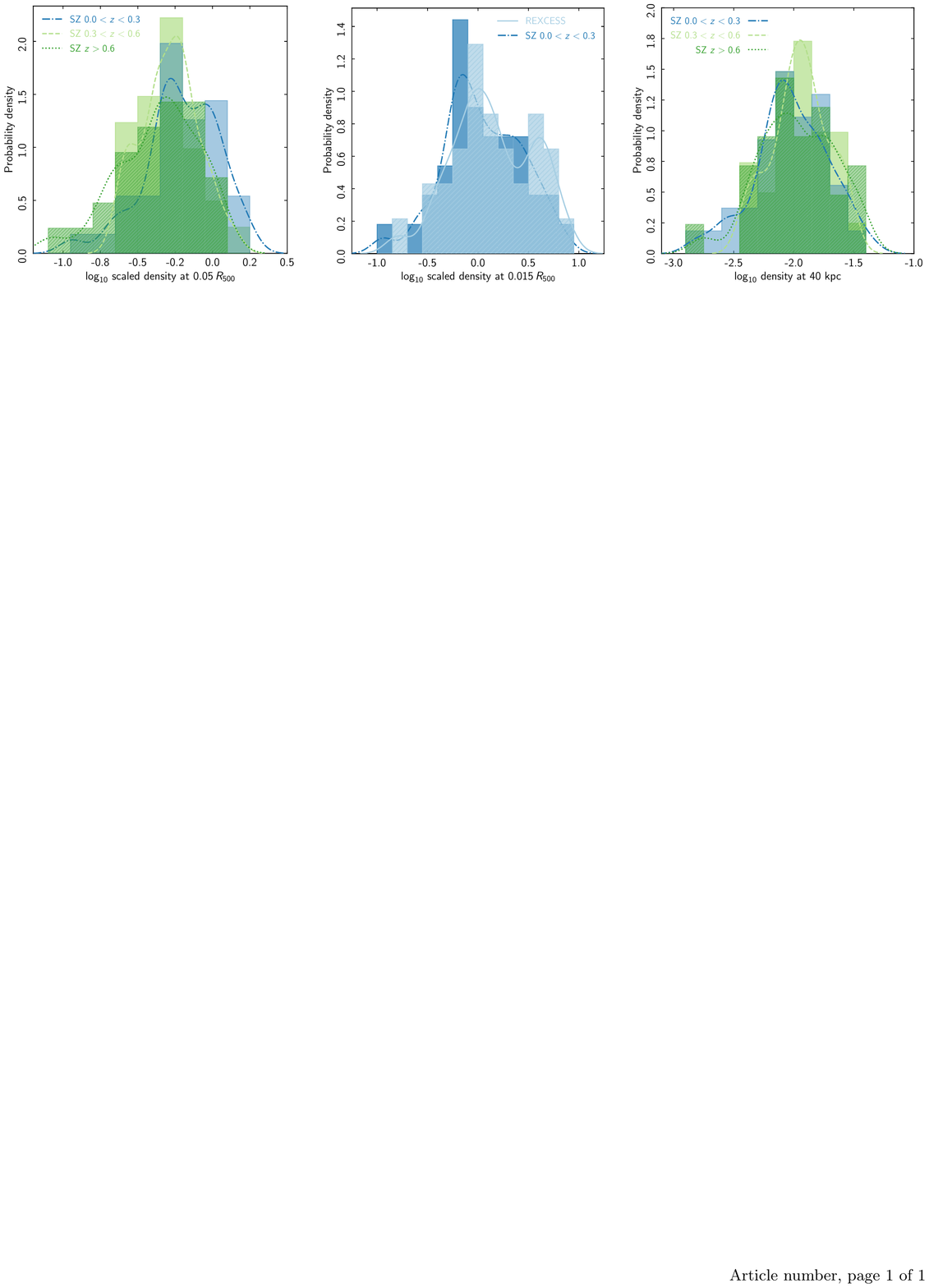}\\
\end{center}
\caption{\footnotesize Central density. {\it Left:} Histogram of central densities for the SZE-selected systems, scaled according to the best-fitting model (Eqn.~\ref{eqn:bestmod1}) derived in Sect.~\ref{sec:modres},  measured at $0.05\,\Rv$. The solid line is a kernel density plot with a smoothing width of 0.15. {\it Middle:} Histogram of scaled central densities for the $z<0.3$ SZE-selected systems compared to the X-ray-selected sample, measured at $0.015\,\Rv$. The solid line is  a kernel density plot with a smoothing width of 0.15. {\it Right:} Histogram of central densities for the SZE-selected sample at 40\,kpc.
}
\label{fig:centdens}
\end{figure*}

Within $R \lesssim 0.2\Rv$ the intrinsic scatter in the scaled gas density profiles climbs steeply towards the centre. This increase is intimately linked to the complex physics of the core regions, dynamical activity, and to the presence or absence of cool core systems in the various samples. In this connection, the sample with the largest intrinsic scatter in the central regions is \rexcess\, reflecting the presence of cool core systems in this dataset.

Beyond $\sim 0.2\Rv$ the relative dispersion of all samples dips below $20\%$, and at $\Rv$ the dispersion in profiles is $\sim 15\%$. In the SZE-selected sub-samples, there is a clear evolution, in the sense that the low-redshift systems exhibit the lowest intrinsic scatter values while the high-redshift systems show higher values. The scatter will be related to intrinsic cluster-to-cluster variations linked to inhomogeneities that will depend on the mass accretion rate and associated dynamical state, together with a component due to uncertainties in the total mass. It is possible that both of these effects conspire to produce higher intrinsic scatter values for higher-redshift systems: one expects an increase in dynamical activity with redshift, while uncertainties in the cluster mass measurement will also increase in the same sense.
%


\subsubsection{Suppression of scatter due to covariance between $\Mgas$ and $\Rv$}\label{sec:dispcov}

The observed scatter in the density profiles may be suppressed by the
use of $\Mgas$ in the computation of $\Rv$ when scaling the radial
coordinate. All other things being equal, a cluster with a higher than
average $\rhog$ (for its mass) at some radius, will have a higher than
average $\Mgas$ and hence $Y_\text{X}$ relative to its mass. Since
$Y_\text{X}$ is then used to estimate $\Rv$ assuming a mean scaling
relation, one would then overestimate $\Rv$ for this cluster. The
radial scaling for this cluster would then be too large, which would
move its density profile back towards the mean profile, reducing the
apparent scatter. The reduction in scatter will depend on the slope of
the density profile (i.e. the reduction is larger where the profile is
steeper), so will be radially dependent.

The possible magnitude of this effect was estimated by generating synthetic
cluster density profiles with a known amount of scatter, and then
scaling them in radius following the method used for the observed
clusters in order to test how much the scatter was changed. In more
detail, the best-fitting median profile presented in Sect.~\ref{sec:modres} was normalised to match a $6~\keV$ cluster at $z=0.15$
(i.e. when integrated to $\Rv$, the gas mass and $Y_{\rm X}$ were
consistent with the scaling relations used for the observed clusters).
For these reference values, the `true' $\Rv$ is $1210$ kpc.

A large number of realisations of this median profile were then
generated by resampling the normalisation from a lognormal
distribution with a standard deviation of $0.2$ (approximating a
constant $20\%$ scatter in $\rhog$ at all radii). For each
realisation, $\Rv$ was then computed in the same manner as for the
observed clusters; the profile was integrated to compute $Y_{\rm X}$ (assuming a
fixed temperature of $6\keV$) and hence $\Rv$, with the process
performed iteratively until $\Rv$ converged. The profile was then
scaled in radius by $\Rv$ and the process was repeated for each
realisation of the density profile.

When the distribution of densities in the realisations was measured at
the `true' value of $\Rv=1210$ kpc, the input scatter of $20\%$ was
recovered. However, when the profiles were each scaled in radius by
the value of $\Rv$ estimated for each realisation from the $M-Y_{\rm X}$ relation, the scatter at a
scaled radius of unity was found to be $10\%$; that is, the scatter is
suppressed by a factor of two at around $\Rv$ due to the dependence of
$\Rv$ on $\Mgas$ in our analysis. The same factor of two suppression was
found for different values of the input scatter. We calculated the reduction in scatter at different scaled radii, obtaining a radial profile of the suppression factor. The intrinsic scatter profile corrected for this suppression factor is plotted in gold in Fig.~\ref{fig:envscatter}. This method makes a number of simplifying assumptions (e.g. there is
no scatter in $T$, the profile is fixed at the median form), but based on this analysis, we estimate that the scatter measured for the observed clusters is likely to be underestimated by a factor of approximately two at $\Rv$.


\subsubsection{Change of central regions over time}\label{sec:centreg}

\citet{mcd17} showed that while the ICM outside the core regions of their SZE-selected sample, covering the redshift range $0.25 < z < 1.2$, evolved self-similarly with redshift,  the central absolute median density (i.e. expressed in units of cm$^{-3}$) did not. They interpreted this result as being due to an un-evolving core component embedded in a self-similarly evolving bulk. 

Our sample is of comparable size to that of \citet{mcd17}, but the evolution with mass and redshift has been decoupled and quantified (Sect.~\ref{sec:gasmodel}).  The effective $5^{\prime\prime}$ resolution of \xmm\ after PSF correction \citep{bar17} allows us to measure the density of the SZE-selected sample across all redshifts down to a scaled radius of $R\sim0.05\,\Rv$. The left-hand panel of  Fig.~\ref{fig:centdens} shows a histogram of the resulting values\footnote{The measurement requires a small extrapolation, by $<1^{\prime\prime}$ in the log-log plane, for seven systems. } scaled according to the best-fitting model (Eqn.~\ref{eqn:bestmod1}) established in Sect.~\ref{sec:modres}. The SZE-selected sample was further divided into three redshift bins to better visualise how the sample changes over time. This histogram is characterised by a strong peak centred on a scaled central density of $-0.2$ (in log space), which is clearly visible, and coincident, in all three SZE-selected sub-samples. While the histogram of the $z>0.6$ sub-sample has no detectable skewness, the histogram of the $z<0.3$ sub-sample exhibits a distinct tail to higher scaled central densities, which is characterised by a moderately large positive skewness of $G_1 = 0.86$ that is significant at $>90\%$ \citep{doa11}. This may indicate the gradual appearance of objects with more peaked scaled central densities towards lower redshifts.

At $z<0.3$, the effective $5^{\prime\prime}$ resolution of \xmm\ after PSF correction allows us to measure the density down to a scaled radius of $R\sim0.015\,\Rv$. The middle  panel of Fig.~\ref{fig:centdens} shows a histogram of the scaled central density of the SZE-selected clusters at $z<0.3$ compared to that of \rexcess. The positive skewness of the SZE-selected systems is confirmed at greater significance ($G_1 = 1.06$), while the histogram of the \rexcess\ sample exhibits two peaks in scaled central density: a main peak that is coincident with the peak of the SZE-selected sub-samples, and a secondary peak at a scaled density of $~0.7$ (in log space). The latter peak is due to cool core systems and may indicate that centrally peaked systems are over-represented in X-ray-selected samples, as has been argued by  \citet{ros17} from their comparison of the image concentration parameter in \planck\ clusters to those for X-ray-selected systems, and also by \citet{and17}. 

The right-hand panel of Fig.~\ref{fig:centdens} shows the histogram of the central density of the SZE-selected sample at 40~kpc, measured in physical units (cm$^{-3}$). There is a broad maximum at $n_{\rm e, 40\,kpc}  \sim 0.01\ {\rm cm}^{-3}$, and the histograms of the three sub-samples coincide. A Kolmogorov-Smirnov test indicates that all three sub-samples come from the same parent distribution. This result suggests, in agreement with \citet{mcd17}, that the absolute central density remains constant over the redshift range probed by the current sample.


\subsection{The \Lxc$-M$  relations}

\subsubsection{Comparison with other work}

\begin{figure}[!t]
\begin{center}
\includegraphics[scale=1.,angle=0,keepaspectratio,width=0.975\columnwidth]{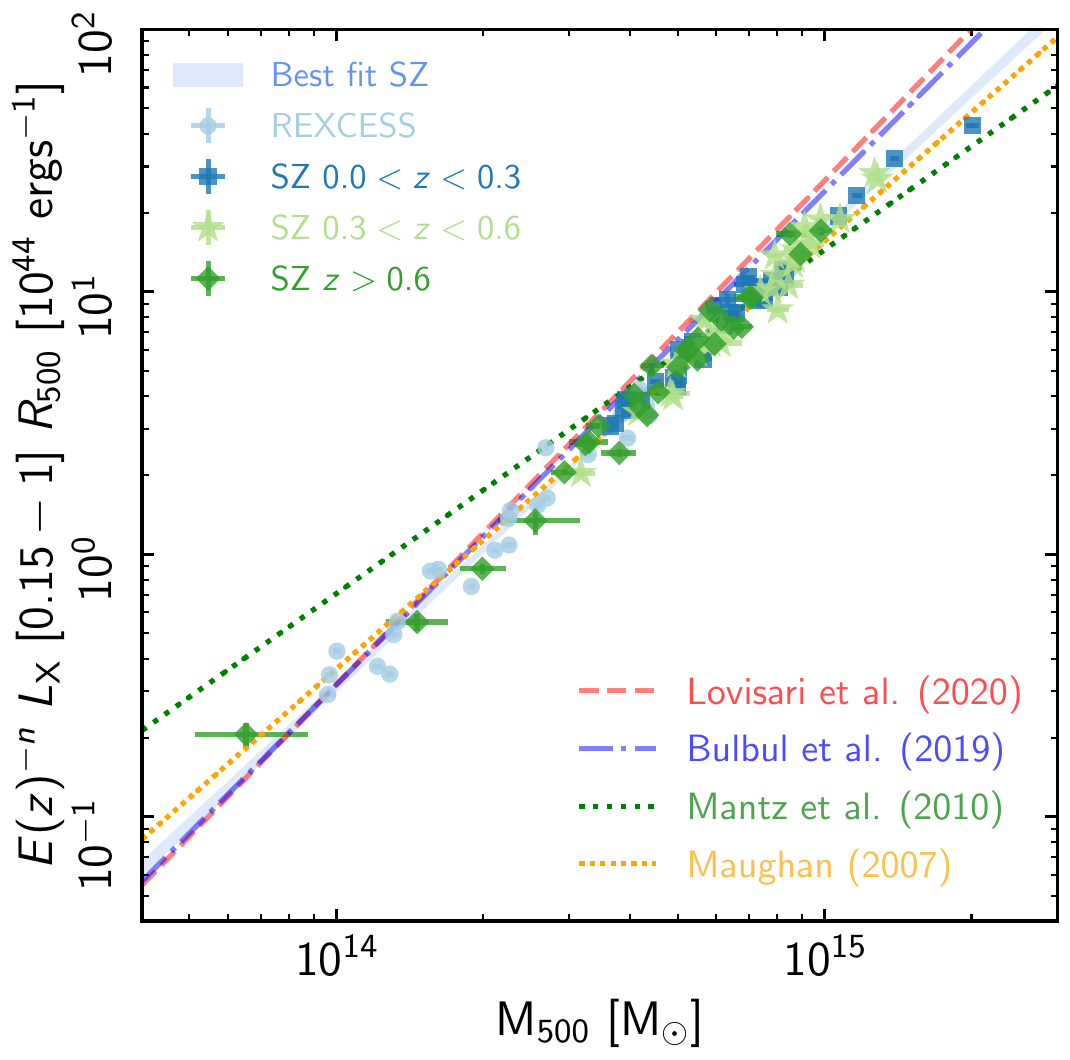}
\end{center}
\caption{\footnotesize Comparison of the bolometric \Lxc$ - M$ relation to previous work \citep{mau07,man10,bul19,lov20}.
}
\label{fig:lmcomp}
\end{figure}

Figure~\ref{fig:lmcomp} shows the best-fitting bolometric \Lxc$-M$ relation for the SZE-selected clusters in the present sample compared to a number of results from the literature \citep{mau07,man10,bul19,lov20}.  With the exception of those obtained by \citet{man10}, these  studies generally find slopes that are steeper than the self-similar expectation of $4/3$, ranging from $1.63$ to $1.92$.  Studies that put constraints on the evolution with redshift \citep{man10,bul19,lov20} generally find good agreement with self-similar expectations (although with large uncertainties). However, any measurement of the dependence of a quantity on the mass will be affected strongly by the sample selection and on how the mass itself has been measured. Data fidelity and sample sizes are now such that systematic effects are starting to become dominant over measurement uncertainties. 

Concerning the sample selection, the results in Sect.~\ref{sec:lxmres} show that the \Lxc$-M$ relations of X-ray- and SZE-selected systems are in good agreement, suggesting that once the core regions are excluded, effects due to detection methods relying on the ICM do not have any impact. Similarly, \citet{lov20} showed that their relaxed and disturbed samples had similar \Lxc$-M$  relation slopes and normalisations.  This suggests that the \Lxc$-M$  relation may also be relatively robust to selection effects linked to cluster dynamical state, likely due to the small intrinsic scatter.

A more fundamental issue is the mass measurement itself \citep{pra19}. In the present work we have used $Y_{\rm X}$ as a mass proxy; the works listed above use variously $Y_{\rm X}$ \citep{mau07}, the gas mass \citep{man10}, the SZE signal-to-noise \citep{bul19}, and the hydrostatic mass \citep{lov20}, as proxies. In this context, the shallower slope of the \citet{man10} relation compared to the others can be fully explained by their assumption of a constant gas mass fraction in the mass calculation \citep[e.g.][]{roz14}.

All of the above mass estimates are derived from ICM observables, and all except \citet{lov20} use scaling laws that have been calibrated on X-ray hydrostatic mass estimates. Independent mass measurements, such as those available from lensing, galaxy velocity dispersions, or caustic measurements \citep[e.g.][]{mau16}, are critical to making progress on this issue. In this connection, weak-lensing mass measurements for individual clusters have been carried out by several projects, such as the Local Cluster Substructure Survey
\citep[LoCuSS;][]{Okabe10b,Okabe13,Okabe16b},  the Canadian Cluster Comparison Project \citep[CCCP;][]{Hoekstra12,Hoekstra15}, the Cluster Lensing And Supernova survey with Hubble
\citep[CLASH;][]{Merten15,Umetsu14,Umetsu16}, Weighing the Giants
\citep[WtG;]{vonderLinden14a,Kelly14,Applegate14}, and CHEX-MATE \citep{CM21}. The cluster community is undertaking a major ongoing effort to critically compare various mass estimates, obtained from mass proxies and from direct X-ray, lensing, or velocity dispersion analyses   \citep[e.g.][]{roz14b, ser15a,ser15b,gro16,ser17}. Ultimately, this effort will help to better constrain the parameters of the scaling relations.

\subsubsection{Link between density profile and \Lxc}

As noted in for example \citet[][]{mau08}, the similarity of the ICM density profiles outside the core implies a low scatter in \Lxc, as is indeed observed here. In order to explore how much of the scatter in \Lxc\ is due to the variation in density profiles, we computed a `pseudo luminosity' $\Lxcp  =\int \rhog^2 \TX^{1/2} \dif V$ for each cluster. For this calculation, we used the measured density profile for each cluster and assumed isothermality at the measured core-excised temperature for the cluster. The integral was performed over a cylindrical volume from projected radii of $0.15\Rv$ to $\Rv$. For 16 of 118 clusters, the density profiles did not reach $\Rv$, so the integrals were truncated at the maximum observed radius. In all cases the profiles reached to $\approx 90\%$ of $\Rv$, and the contribution to the luminosity of the outer parts of the profile is very small, so the effect of this truncation is negligible. The scatter in $\Lxcp$ then provides an estimate of the scatter in the bolometric \Lxc\ due only to the scatter in density profiles.

The intrinsic scatter about the best fitting relation to $\Lxcp\, E(z)^{-2.5}$ versus $\Mv$ was then measured (assuming that the fractional statistical error on $\Lxcp$ is the same as that in \Lxc\ for each cluster), giving a value of $11\%$. This implies that most or all of the intrinsic scatter in the bolometric \Lxc\ at fixed mass can be explained by the variation in the ICM density profiles. 

In principle, the use of $M_{\rm Yx}$ to determine $\Rv$ and hence the aperture within which \Lxc\ and $\Lxcp$ are measured, could introduce additional scatter in the \Lxc$-\Mv$ relation. If $M_{\rm Yx}$ were scattered high relative to the true mass of a cluster, then $\Rv$ would be overestimated and the $0.15-1\Rv$ aperture would be shifted to larger radii. This shift would reduce \Lxc\ since more of the luminosity comes from the inner edge of the aperture than the outer edge. Hence, if $M_{\rm Yx}$ were scattered high, then \Lxc\ and $\Lxcp$ would be scattered low, adding to the observed scatter in the $\Lxcp-\Mv$ relation. We examined the impact of this effect by adding $10\%$ scatter to $M_{\rm Yx}$ when computing $\Lxcp$. This increased the measured scatter in $\Lxcp$ by less than $1\%$. The dependence of the luminosity aperture on $M_{\rm Yx}$ leads to a negligible contribution to the scatter in the \Lxc$-M_{\rm Yx}$ relation.

We therefore conclude that the measured intrinsic scatter in the bolometric \Lxc\ at fixed mass is dominated almost entirely by the variation in the ICM density profiles. The results do not depend on the aperture for reasonable assumptions on the scatter between $M_{\rm Yx}$ and the true mass $\Mv$. Any residual scatter will come from inhomogeneity and/or substructure 
in the density distribution, or from the effects of structure in the temperature or metallicity distribution.


\subsubsection{Impact of selection bias and covariance}\label{sec:seleff}

We found above that the bolometric \Lxc$-M$ relation has a slope $B_{\rm L} \sim1.7$, which is steeper than the self-similar value of 1.3, and has a small scatter of $\sigma_{\ln{L_{\rm Xc}}} \sim 0.15$. A similar difference in slope of $\Delta\,B_{\rm L} \sim +0.4$ is seen relative to the self-similar soft-band \Lxc$-M$ relation. One might wonder whether and to what extent selection effects and covariance may impact these results. Firstly, in survey data the observable that is used for cluster detection, $\mathcal{O}_{\rm det}$, can be affected by so-called Malmquist bias. This happens because clusters that are scattered to higher values of $\mathcal{O}_{\rm det}$, whether by noise or by the intrinsic scatter between the observable and the mass, will be preferentially detected, leading to a positive bias in the average value of $\mathcal{O}_{\rm det}$. This is a particular concern near the detection threshold, and will affect the apparent slope of any scaling relation with $\mathcal{O}_{\rm det}$. Secondly, the intrinsic scatter of each quantity around the mean relation may well be correlated, leading to a reduction in the measured scatter with respect to the true underlying value.

Since we showed in Sect.~\ref{sec:lxmres} that the results for the X-ray-selected sample are compatible within $1\sigma$ with those for the local SZE-selected sample, we focus here on the SZE selection. As detailed in Sect.~\ref{sec:dataset}, the SZE-selected sample is composed of four different sub-samples and with different selections: the ESZ, LP1, LP2 and LP3.  Although the ESZ sub-sample was generated from high signal-to-noise ratio (${\rm S/N} > 6$) detections, \cite{fas21} showed that its selection is dominated by the instrumental and astrophysical scatter, and that the intrinsic scatter in the $Y_{\rm SZ}$-$M$ relation has a negligible effect on the recovered scaling parameters.
The three other SZE sub-samples (LP1, LP2, LP3) are selected at an S/N lower than that of the  ESZ sample, and are thus even more affected by the instrumental and astrophysical scatter. We restricted in consequence our selection study to the ESZ. This will provide upper limits to the effects of selection on the results for the \Lxc$-M$ of the SZE-selected systems in the current study.

Appendix~\ref{appx:sims} describes in detail the simulations we used, which were similar to those produced for the \citet{fas21} study. Simulated clusters, modelled with the \citet{arn10} pressure profile and drawn from a Tinker mass function \citep{tin08}, were injected into the Planck Early SZ maps in the `cosmological' mask region. The Multi-Matched Filter extraction algorithm \citep{mel06} was then applied, to obtain SZ detections at ${\rm S/N} > 6$, corresponding to the threshold for the ESZ sample.  We then matched the injected and recovered clusters to produce a mock ESZ catalogue, doing this twenty times to generate 3188 detections in total. To investigate selection and covariance effects, we assumed a Gaussian lognormal correlated distribution for $Y_{\rm SZ}$, $Y_{\rm X}$, and \Lxc\ at fixed true mass $M$, with a covariance matrix to describe correlations between parameters. 

A first-order estimate of the expected scatter in the \Lxc$-M_{\rm Yx}$ relation can be obtained by assuming that selection effects are negligible. In this case, 
\begin{gather}
\sigma^2_{\ln{L_{\rm Xc}|M_{\rm Yx}}} = \sigma^2_{\ln{L_{\rm Xc}|M}} + \left({B_{\rm L}^2 / B_{\rm Y}^2} \right)\, \sigma^2_{\ln{Y_{\rm X}|M}}-2\,t\,\sigma_{\ln{L_{\rm Xc}|M}}\, \sigma_{\ln{Y_{\rm X}|M}}.\label{eqn:forder}
\end{gather}
For a measured intrinsic scatter $\sigma_{\ln{L_{\rm Xc}|M_{\rm Yx}}} \sim 0.13$, a covariance between \Lxc\ and $Y_{\rm X}$ of $t = 0.85$ \citep{af19}, and assuming a scatter between $Y_{\rm X}$ and true mass $M$ of $\sigma_{\ln{Y_{\rm X}|M}} \sim 0.16$ motivated by recent numerical simulations \citep{sp14,alb17,nt18}, the resulting dispersion between core-excised luminosity and true mass is $\sigma_{\ln{L_{\rm Xc}|M}} \sim 0.22$. This suggests that the measured scatter is underestimated by a factor of $0.22/0.13 \sim 1.7$ due to the use of $Y_{\rm X}$ as a mass proxy. 

We generated a series of simulations using the above values for true scatter and covariance between quantities, and used them to investigate the effects of selection, intrinsic scatter, and covariance on the results. We found:

\begin{itemize}

\item The Malmquist bias induced on the \Lxc$-Y_{\rm SZ}$ relation due to intrinsic scatter in the $D_{\rm A}^2\,Y_{\rm SZ}-M$ relation is completely negligible, with the slope changing by less than 0.5\%. The resulting impact on the slope of the \Lxc$-M$ relation is $\Delta B_{\rm L} = -0.01$. This is important, because it implies that selection effects due to intrinsic scatter in the $Y_{\rm SZ}$ observable cannot account for the observed steeper slope of the \Lxc$-M$ relations seen here.

\item Use of $Y_{\rm X}$ as a mass proxy introduces additional intrinsic scatter with respect to the underlying mass. The net impact on the recovered slope of the \Lxc$-M$ relation is minimal, with $\Delta B_{\rm L} = -0.04$.

\item Intrinsic scatter in the \Lxc$-Y_{\rm X}$ relation has the effect of redressing the slope towards its original value, with $\Delta B_{\rm L} = +0.02$. 

\item Covariance between $Y_{\rm X}$ and \Lxc\ again changes the slope. For $t=0.85$ \citep{af19}, the impact on the slope of the \Lxc$-M$ relation is $\Delta B_{\rm L} = +0.04$.

\end{itemize}

We thus conclude that the slope of the \Lxc$-M_{\rm Yx}$ relation found here is robust to selection effects due to intrinsic scatter in the $Y_{\rm SZ}$ and $Y_{\rm X}$ proxies, and to covariance between quantities. The dispersion, however, is very sensitive to the covariance. We estimate that the measured dispersion in the relation between the core-excised luminosity and true mass,  \Lxc$-M$, is underestimated by a factor of $\sim 1.7$ due to the use of $Y_{\rm X}$ as a mass proxy.

\subsection{Link to $f_{\rm gas} -M$}

The \Lxc$-M$ relations derived above have a steeper dependence than self-similar, which cannot be explained by selection effects, intrinsic scatter, or covariance. There is evidence for a dependence of the gas content with mass, which has the effect of suppressing the luminosity preferentially in lower-mass systems, leading to the observed steepening of the relation \citep[see discussion in][]{pra09}.  Interestingly, with the assumption of a standard dependence of temperature on mass ($T \propto M^{2/3}$, e.g. \citealt{arn05,man16}), use of Eqn.~\ref{eqn:lxeqn} with the observed bolometric \Lxc$\propto M^{1.7}$ relation yields $f_{\rm gas} \propto M^{0.21}$.

The gas density profile model detailed in Sect.~\ref{sec:gasmodel} yields an alternative method to obtain the dependence of the gas mass fraction with mass, as the model includes a dependence of $f_{\rm gas} = M_{\rm gas} / M$ on the total mass $M$, via Eqn.~\ref{eqn:fgasd}. The  best-fitting gas density model (Eqn~\ref{eqn:bestmod1}) suggests $f_{\rm gas} \propto M^{0.22\pm0.01}$. This dependence of the gas content on total mass is therefore in good agreement with the expected relation given the observed \Lxc$-M$ dependence, and with previous findings \citep[e.g.][]{pra09,lov15,ett15} based primarily on X-ray-selected clusters.


\section{Conclusions}

We have examined the gas density profiles and the relation between the core excised X-ray luminosity \Lxc\ and the total mass derived from the $Y_{\rm X}$ mass proxy for 118 X-ray and SZE-selected objects covering a mass range of $M_{500} = [0.5-20] \times 10^{14}\,$~M$_{\odot}$ and extending in redshift up to $z\sim 1.13$. We first examined the scaled density profiles:
\begin{itemize}
\item The gas density profiles do not scale perfectly self-similarly, exhibiting subtle trends in mass and redshift. 

\item Motivated by this finding, we fitted an analytic gas density model to the 93 SZE-selected systems. The analytic model is based on a generalised NFW profile, and correctly reproduces the scaled gas density profile and the radial variation of its intrinsic dispersion. Combined with the empirical mass scaling of the profiles, this analytic model defines the gas density profile of SZE-selected clusters as a function of mass and redshift. This model is given in Eqns.~\ref{eqn:fullmod}-\ref{eqn:bestmod2}.

\item The intrinsic dispersion in scaled profiles is greatest in the central regions, declining to a minimum at $\sim 0.5-0.7\,\Rv$, and increasing thereafter. The dispersion is similar for X-ray-selected clusters and for local SZE-selected clusters, except in the centre, where the X-ray-selected systems have a higher dispersion. There is a hint for an evolution of the dispersion with redshift, which may be linked to an increase in perturbed clusters at higher redshifts.

\item We investigated the effect of covariance between $M_{\rm gas}$ and $\Rv$ due to the use of $M_{\rm Y_{\rm X}}$ as a mass proxy, obtaining a radial profile of the scatter suppression factor. Taking into account this suppression factor, we estimated a scatter in scaled density profiles of approximately $40\%$ at $\Rv$.

\item We quantified deviations from the average scaling with radius. These show no variation with mass, but which show a significant variation with redshift, in the sense that the core regions clearly evolve differently as compared to the bulk.

\item We examined the scaled central density measured at $R = 0.05\,\Rv$ for the SZE-selected systems, finding that only the $z<0.3$ sample is skewed. This skewness is positive, and  may indicate the increased presence of centrally peaked systems at later times.

\item We measured the scaled central density at $R = 0.015\,\Rv$ for the X-ray and SZE-selected systems at $z<0.3$. The scaled central density of the local X-ray-selected sample exhibits two peaks. The main peak, corresponding to non-cool core systems in the X-ray-selected sample, is slightly offset to higher scaled central density from that of the local SZE-selected sample. The secondary peak in the X-ray-selected sample, corresponding to the cool core systems, is not seen in the SZE-selected sample, although the latter does exhibit a clear tail to higher scaled central density as confirmed by the strongly positively skewed distribution. 

\item The absolute value of the central density in the SZE-selected sample measured at 40~kpc does not appear to evolve with redshift, consistent with the findings of \citet{mcd17}.

\end{itemize}

We then examined the relation between the  core excised X-ray luminosity \Lxc\ and the total mass derived from the $\YX$ mass proxy, $M_{\rm Yx}$. 

\begin{itemize}

\item This relation is extremely tight, with a logarithmic intrinsic scatter of $\sigma_{\ln{L_{\rm xc}|M_{\rm Yx}}} \lesssim 0.15$ depending on sub-sample and band in which the luminosity is measured. Importantly, at low redshift, the best-fitting parameters of this relation  do not depend on whether the sample was selected in X-rays or through the SZE, suggesting that \Lxc\ is a selection-independent quantity. 

\item The slope of the bolometric relation fitted to the SZE-selected clusters, $B \sim 1.74\pm0.02$, is significantly steeper than self-similar. When left free to vary, the evolution of $n = -2.5\pm0.09$ is in agreement with the self-similar value of $-7/3$ within $< 2\sigma$. 

\item We thoroughly examined the impact of selection bias and covariance on the relation. We found that the slope of the \Lxc$-M_{\rm Yx}$ relation is robust to selection effects due to intrinsic scatter in the $Y_{\rm SZ}$ and $Y_{\rm X}$ proxies, and to covariance between quantities. The dispersion, however, is very sensitive to the covariance. For reasonable values of covariance, we  estimate that the measured dispersion in the \Lxc$-M$ relation is underestimated by a factor of at most $\sim 1.7$ due to the use of $Y_{\rm X}$ as a mass proxy, implying a true scatter of $\sigma_{\ln{L_{\rm Xc}|M}} \sim 0.22$.

\item We show explicitly that the scatter in the \Lxc$-M$ relation can be accounted for almost entirely by object-to-object variations in gas density profiles. 

\end{itemize}

With our study we have examined the mass and redshift dependence of the ICM gas density profile, and made quantitative comparisons between X-ray- and SZE-selected samples. Our overall conclusion is consistent with the view that the ICM bulk evolves approximately self-similarly, with the core regions evolving separately due to cooling and feedback from the central active galactic nucleus. Indeed, it suggests potentially subtle differences in the core regions between X-ray- and SZE-selected systems. It also supports a view where the ICM gas mass fraction depends on mass up to high redshift, with a dependence $f_{\rm gas} \propto M_{500}^{0.22\pm0.02}$ for the present sample. Further progress can be undoubtedly be made by bringing to bear fully independent mass estimates, such as those that can be obtained from weak lensing and/or galaxy velocity dispersions. Such studies are one of the goals of the CHEX-MATE project \citep{CM21}.

\begin{acknowledgements}
GWP, MA, and J-BM acknowledge funding from the European Research  Council  under  the  European  Union’s  Seventh  Framework Programme (FP72007-2013) ERC grant agreement no. 340519, and from the French space agency, CNES. BJM acknowledges support from the Science and Technology Facilities Council (grant number ST/V000454/1). The results reported in this article are based on data obtained from the XMM-Newton observatory, an ESA science mission with instruments and contributions directly funded by ESA Member States and NASA.

\end{acknowledgements}

\bibliographystyle{aa}
\bibliography{lmfinalc}


\longtab{
\begin{longtable}{lrrrrrrrr}
\caption{\label{Data} Sample observational data.}\\
\toprule
\toprule
\vspace{-3mm}\\
Planck name & \rexcess\ name & $z$ & RA & Dec. & $M_{500}$ &  \Lxc [0.5-2] keV & \Lxc bol\\
 &  &  & [deg] & [deg] & $[10^{14}$ M$_{\odot}]$ &  [$10^{44}$ erg s$^{-1}$]  &  [$10^{44}$ erg s$^{-1}$]\\
\vspace{-3mm}\\
\midrule
\endfirsthead
\vspace{-3mm}\\
\caption{continued.}\\
\midrule
\midrule
\vspace{-3mm}\\
Plank name & \rexcess\  name & $z$ & RA & Dec. & $M_{500}$ &  \Lxc [0.5-2] keV & \Lxc bol\\
\vspace{-3mm}\\
 &  &  & [deg] & [deg] & $[10^{14}$ M$_{\odot}]$ &  [$10^{44}$ erg s$^{-1}$]  &  [$10^{44}$ erg s$^{-1}$]\\

\midrule
\vspace{-3mm}\\
\endhead
\bottomrule
\vspace{-2mm}\\
\endfoot
&         RXC\,J$2023.0-2056$  & $0.056    $  & $305.7450    $  & $-20.9485    $& $ 1.21^{+0.03}_{-0.03}                        $& $ 0.16^{+0.00}_{-0.00} $& $ 0.40^{+0.01}_{-0.01} $ \\
&         RXC\,J$2157.4-0747$  & $0.058    $  & $329.3673    $  & $ -7.8046    $& $ 1.29^{+0.03}_{-0.03}                        $& $ 0.15^{+0.00}_{-0.00} $& $ 0.37^{+0.01}_{-0.01} $ \\
&         RXC\,J$0345.7-4112$  & $0.060    $  & $ 56.4428    $  & $-41.2042    $& $ 0.97^{+0.02}_{-0.02}                        $& $ 0.15^{+0.00}_{-0.00} $& $ 0.37^{+0.01}_{-0.01} $ \\
&          RXC\,J$0225.1-2928$  & $0.060    $  & $ 36.2877    $  & $-29.4773    $& $ 0.96^{+0.04}_{-0.04}                       $& $ 0.12^{+0.00}_{-0.00} $& $ 0.31^{+0.01}_{-0.01} $ \\
&          RXC\,J$1236.7-3354$  & $0.080    $  & $189.1712    $  & $-33.9260    $& $ 1.33^{+0.02}_{-0.02}                       $& $ 0.24^{+0.00}_{-0.00} $& $ 0.61^{+0.01}_{-0.01} $ \\
&          RXC\,J$2129.8-5048$ & $0.080    $  & $322.4271    $  & $-50.8167    $& $ 2.26^{+0.06}_{-0.06}                        $& $ 0.41^{+0.01}_{-0.01} $& $ 1.19^{+0.02}_{-0.02} $ \\
 PSZ2 G093.92+34.92 &                & $0.081    $  & $258.2029    $  & $ 64.0636    $& $ 5.00^{+0.21}_{-0.15}                  $& $ 1.24^{+0.01}_{-0.01} $& $ 4.78^{+0.10}_{-0.10} $ \\
PSZ2\,G222.52+20.58 &         RXC\,J$0821.8+0112$  & $0.082    $  & $125.4614    $  & $  1.1967    $& $ 1.31^{+0.03}_{-0.04}    $& $ 0.22^{+0.00}_{-0.00} $& $ 0.54^{+0.01}_{-0.01} $ \\
PSZ2 G306.77+58.61 &             & $0.084    $  & $194.8438    $  & $ -4.1983    $& $ 4.51^{+0.12}_{-0.11}                      $& $ 1.48^{+0.01}_{-0.01} $& $ 5.02^{+0.07}_{-0.07} $ \\
PSZ2 G306.66+61.06 &              & $0.084    $  & $194.6734    $  & $ -1.7622    $& $ 4.21^{+0.05}_{-0.05}                     $& $ 1.27^{+0.00}_{-0.00} $& $ 4.25^{+0.03}_{-0.03} $ \\
PSZ2\,G308.64+60.26 &          RXC\,J$1302.8-0230$  & $0.085    $  & $195.7216    $  & $ -2.5170    $& $ 1.89^{+0.03}_{-0.03}   $& $ 0.31^{+0.00}_{-0.00} $& $ 0.83^{+0.01}_{-0.01} $ \\
PSZ2\,G099.57-58.64 &          RXC\,J$0003.8+0203$  & $0.092    $  & $  0.9572    $  & $  2.0657    $& $ 2.11^{+0.04}_{-0.04}   $& $ 0.40^{+0.00}_{-0.00} $& $ 1.15^{+0.01}_{-0.01} $ \\
PSZ2 G321.98-47.96 &             & $0.094    $  & $342.4917    $  & $-64.4294    $& $ 3.73^{+0.06}_{-0.06}                      $& $ 1.07^{+0.00}_{-0.00} $& $ 3.51^{+0.03}_{-0.03} $ \\
PSZ2 G336.60-55.43 &             & $0.097    $  & $341.5711    $  & $-52.7261    $& $ 3.99^{+0.08}_{-0.08}                      $& $ 1.21^{+0.01}_{-0.01} $& $ 3.95^{+0.04}_{-0.04} $ \\
PSZ2\,G311.62-42.31  &         RXC\,J$2319.6-7313$  & $0.098    $  & $349.9167    $  & $-73.2277    $& $ 1.56^{+0.03}_{-0.03}   $& $ 0.38^{+0.01}_{-0.01} $& $ 0.97^{+0.02}_{-0.01} $ \\
PSZ2\,G332.23-46.37 &          & $0.099    $  & $330.4720    $  & $-59.9454    $& $ 5.64^{+0.09}_{-0.09}                        $& $ 1.76^{+0.01}_{-0.01} $& $ 6.61^{+0.06}_{-0.06} $ \\
&          RXC\,J$0211.4-4017$  & $0.101    $  & $ 32.8528    $  & $-40.2915    $& $ 1.00^{+0.02}_{-0.02}                       $& $ 0.20^{+0.00}_{-0.00} $& $ 0.48^{+0.01}_{-0.00} $ \\
&          RXC\,J$0049.4-2931$  & $0.108    $  & $ 12.3460    $  & $-29.5206    $& $ 1.62^{+0.04}_{-0.04}                       $& $ 0.38^{+0.01}_{-0.01} $& $ 1.00^{+0.02}_{-0.02} $ \\
PSZ2 G053.53+59.52 &              & $0.113    $  & $227.5528    $  & $ 33.5104    $& $ 5.64^{+0.14}_{-0.14}                     $& $ 1.71^{+0.01}_{-0.01} $& $ 6.34^{+0.10}_{-0.10} $ \\
PSZ2\,G352.28-77.66   &          RXC\,J$0006.0-3443$  & $0.115    $  & $  1.5015    $  & $-34.7224    $& $ 3.95^{+0.12}_{-0.12} $& $ 0.97^{+0.01}_{-0.01} $& $ 3.17^{+0.05}_{-0.05} $ \\
PSZ2\,G255.64-25.30   &         RXC\,J$0616.8-4748$  & $0.116    $  & $ 94.2158    $  & $-47.7950    $& $ 2.70^{+0.05}_{-0.05}  $& $ 0.60^{+0.01}_{-0.01} $& $ 1.88^{+0.02}_{-0.02} $ \\
 PSZ2\,G285.52-62.23   &         RXC\,J$0145.0-5300$  & $0.117    $  & $ 26.2433    $  & $-53.0208    $& $ 4.37^{+0.08}_{-0.08} $& $ 1.08^{+0.01}_{-0.01} $& $ 3.87^{+0.03}_{-0.03} $ \\
&          RXC\,J$1516.3+0005$  & $0.118    $  & $229.0747    $  & $  0.0893    $& $ 3.28^{+0.07}_{-0.04}                       $& $ 0.89^{+0.01}_{-0.01} $& $ 2.76^{+0.02}_{-0.02} $ \\
&          RXC\,J$2149.1-3041$  & $0.118    $  & $327.2817    $  & $-30.7013    $& $ 2.25^{+0.03}_{-0.03}                       $& $ 0.55^{+0.00}_{-0.00} $& $ 1.58^{+0.02}_{-0.01} $ \\
PSZ2\, G277.38+47.07 &          RXC\,J$1141.4-1216$  & $0.119    $  & $175.3513    $  & $-12.2776    $& $ 2.27^{+0.02}_{-0.02}  $& $ 0.60^{+0.00}_{-0.00} $& $ 1.69^{+0.01}_{-0.01} $ \\
PSZ2\, G000.04+45.13 &          RXC\,J$1516.5-0056$  & $0.120    $  & $229.1842    $  & $ -0.9696    $& $ 2.59^{+0.05}_{-0.04}  $& $ 0.62^{+0.01}_{-0.01} $& $ 1.77^{+0.02}_{-0.02} $ \\
PSZ2\, G256.38+44.04 &          RXC\,J$1044.5-0704$  & $0.134    $  & $161.1370    $  & $ -7.0687    $& $ 2.69^{+0.02}_{-0.02}  $& $ 1.04^{+0.01}_{-0.01} $& $ 2.99^{+0.02}_{-0.02} $ \\
PSZ2 G241.79-24.01 &          RXC\,J$0605.8-3518$  & $0.139    $  & $ 91.4752    $  & $-35.3023    $& $ 3.89^{+0.09}_{-0.09}    $& $ 1.27^{+0.01}_{-0.01} $& $ 4.23^{+0.16}_{-0.03} $ \\
PSZ2\,G042.81-82.97 &          RXC\,J$0020.7-2542$  & $0.141    $  & $  5.1755    $  & $-25.7080    $& $ 3.84^{+0.06}_{-0.06}   $& $ 1.17^{+0.01}_{-0.01} $& $ 4.05^{+0.04}_{-0.03} $ \\
 PSZ2 G002.77-56.16 &         RXC\,J$2218.6-3853$  & $0.141    $  & $334.6677    $  & $-38.9018    $& $ 4.92^{+0.11}_{-0.11}    $& $ 1.50^{+0.02}_{-0.02} $& $ 5.56^{+0.06}_{-0.06} $ \\
PSZ2 G226.18+76.79 &                & $0.143    $  & $178.8250    $  & $ 23.4049    $& $ 6.24^{+0.06}_{-0.06}                   $& $ 2.27^{+0.01}_{-0.01} $& $ 8.84^{+0.06}_{-0.06} $ \\
PSZ2\,G028.77-33.56 &          RXC\,J$2048.1-1750$  & $0.147    $  & $312.0419    $  & $-17.8413    $& $ 4.31^{+0.07}_{-0.07}   $& $ 1.36^{+0.01}_{-0.01} $& $ 4.40^{+0.03}_{-0.03} $ \\
PSZ2 G236.92-26.65 &          RXC\,J$0547.6-3152$  & $0.148    $  & $ 86.9081    $  & $-31.8727    $& $ 5.01^{+0.10}_{-0.10}    $& $ 1.58^{+0.01}_{-0.01} $& $ 5.73^{+0.38}_{-0.04} $ \\
 PSZ2 G008.47-56.34 &         RXC\,J$2217.7-3543$  & $0.149    $  & $334.4400    $  & $-35.7260    $& $ 3.61^{+0.05}_{-0.05}    $& $ 1.15^{+0.01}_{-0.01} $& $ 3.68^{+0.03}_{-0.03} $ \\
PSZ2 G003.93-59.41 &         RXC\,J$2234.5-3744$  & $0.151    $  & $338.6125    $  & $-37.7360    $& $ 7.36^{+0.09}_{-0.09}     $& $ 3.05^{+0.02}_{-0.02} $& $12.28^{+0.12}_{-0.10} $ \\
 PSZ2 G021.10+33.24 &            & $0.152    $  & $248.1959    $  & $  5.5754    $& $ 8.06^{+0.23}_{-0.22}                      $& $ 2.96^{+0.02}_{-0.02} $& $12.51^{+0.17}_{-0.17} $ \\
PSZ2 G244.71+32.50 & RXC\,J$0945.4 -0839$  & $0.153    $  & $146.3575    $  & $ -8.6557    $& $ 3.91^{+0.15}_{-0.11}        $& $ 1.43^{+0.02}_{-0.02} $& $ 4.66^{+0.08}_{-0.08} $ \\
 PSZ2\,G018.32-28.50 &         RXC\,J$2014.8-2430$  & $0.154    $  & $303.7154    $  & $-24.5059    $& $ 5.38^{+0.07}_{-0.07}   $& $ 2.08^{+0.02}_{-0.02} $& $ 7.47^{+0.07}_{-0.06} $ \\
PSZ2 G049.22+30.87 &        & $0.164    $  & $260.0417    $  & $ 26.6250    $& $ 5.30^{+0.11}_{-0.11}                           $& $ 2.06^{+0.02}_{-0.02} $& $ 7.49^{+0.08}_{-0.08} $ \\
PSZ2 G263.68-22.55 &         RXC\,J$0645.4-5413$  & $0.164    $  & $101.3712    $  & $-54.2273    $& $ 7.38^{+0.18}_{-0.18}     $& $ 2.81^{+0.02}_{-0.02} $& $11.33^{+0.25}_{-0.08} $ \\
PSZ2\, G249.38+33.26 &           RXC\,J$0958.3-1103$  & $0.167    $  & $149.5930    $  & $-11.0644    $& $ 4.17^{+0.22}_{-0.15} $& $ 1.42^{+0.04}_{-0.04} $& $ 5.21^{+0.15}_{-0.14} $ \\
PSZ2 G097.72+38.12 &               & $0.171    $  & $248.9597    $  & $ 66.2125    $& $ 5.37^{+0.10}_{-0.10}                    $& $ 2.20^{+0.01}_{-0.01} $& $ 7.76^{+0.09}_{-0.09} $ \\
PSZ2 G067.17+67.46 &                & $0.171    $  & $216.5105    $  & $ 37.8243    $& $ 8.25^{+0.19}_{-0.19}                   $& $ 3.23^{+0.02}_{-0.02} $& $14.34^{+0.18}_{-0.18} $ \\
 PSZ2 G149.75+34.68 &                & $0.182    $  & $127.7462    $  & $ 65.8398    $& $ 8.07^{+0.53}_{-0.42}                  $& $ 3.04^{+0.06}_{-0.06} $& $13.15^{+0.41}_{-0.41} $ \\
PSZ2 G313.33+61.13 &          RXC\,J$1311.4-0120$  & $0.183    $  & $197.8727    $  & $ -1.3415    $& $ 8.41^{+0.08}_{-0.08}    $& $ 3.36^{+0.01}_{-0.01} $& $14.93^{+0.08}_{-0.07} $ \\
 PSZ2 G195.75-24.32 &            & $0.203    $  & $ 73.5402    $  & $  2.9212    $& $ 8.31^{+0.22}_{-0.21}                      $& $ 3.46^{+0.03}_{-0.03} $& $14.83^{+0.22}_{-0.22} $ \\
PSZ2 G006.76+30.45 &                & $0.203    $  & $243.9399    $  & $ -6.1491    $& $20.10^{+0.61}_{-0.62}            $& $ 9.50^{+0.08}_{-0.08} $& $55.06^{+1.00}_{-1.00} $ \\
 PSZ2 G182.59+55.83 &         & $0.206    $  & $154.2653    $  & $ 39.0482    $& $ 5.02^{+0.10}_{-0.10}                         $& $ 2.15^{+0.02}_{-0.02} $& $ 7.74^{+0.10}_{-0.10} $ \\
 PSZ2 G166.09+43.38 &          & $0.217    $  & $139.4765    $  & $ 51.7315    $& $ 6.61^{+0.14}_{-0.14}                        $& $ 2.75^{+0.02}_{-0.02} $& $10.85^{+0.12}_{-0.12} $ \\
PSZ2 G092.71+73.46 &           & $0.223    $  & $203.8298    $  & $ 41.0001    $& $ 7.38^{+0.18}_{-0.18}                        $& $ 3.58^{+0.02}_{-0.02} $& $13.63^{+0.16}_{-0.16} $ \\
 PSZ2\,G055.59+31.85 &          & $0.224    $  & $260.6127    $  & $ 32.1324    $& $ 6.33^{+0.59}_{-0.51}                       $& $ 3.15^{+0.13}_{-0.13} $& $12.33^{+0.65}_{-0.65} $ \\
  PSZ2 G072.62+41.46 &            & $0.228    $  & $250.0837    $  & $ 46.7107    $& $11.62^{+0.33}_{-0.24}                     $& $ 6.31^{+0.05}_{-0.05} $& $30.96^{+0.54}_{-0.54} $ \\
 PSZ2 G073.97-27.82 &             & $0.231    $  & $328.4031    $  & $ 17.6949    $& $10.67^{+0.28}_{-0.28}                     $& $ 5.61^{+0.05}_{-0.05} $& $25.89^{+0.38}_{-0.38} $ \\
PSZ2 G294.68-37.01 &               & $0.274    $  & $ 45.9366    $  & $-77.8784    $& $ 7.54^{+0.31}_{-0.32}                    $& $ 3.16^{+0.04}_{-0.04} $& $13.35^{+0.28}_{-0.28} $ \\
 PSZ2 G241.76-30.88 &              & $0.275    $  & $ 83.2323    $  & $-37.0270    $& $ 6.40^{+0.23}_{-0.23}                    $& $ 2.96^{+0.04}_{-0.04} $& $11.77^{+0.23}_{-0.23} $ \\
PSZ2 G259.98-63.43 &                & $0.284    $  & $ 38.0772    $  & $-44.3464    $& $ 6.99^{+0.21}_{-0.20}                   $& $ 4.34^{+0.04}_{-0.04} $& $16.28^{+0.27}_{-0.27} $ \\
PSZ2 G244.37-32.15 &                 & $0.284    $  & $ 82.2211    $  & $-39.4714    $& $ 6.86^{+0.30}_{-0.33}                  $& $ 4.18^{+0.05}_{-0.05} $& $15.29^{+0.37}_{-0.37} $ \\
PSZ2 G106.87-83.23 &               & $0.292    $  & $ 10.8519    $  & $-20.6229    $& $ 5.91^{+0.25}_{-0.18}                    $& $ 3.38^{+0.04}_{-0.04} $& $12.36^{+0.25}_{-0.25} $ \\
PSZ2 G262.27-35.38 &                & $0.295    $  & $ 79.1536    $  & $-54.5090    $& $ 7.81^{+0.81}_{-0.73}                   $& $ 3.90^{+0.17}_{-0.17} $& $15.98^{+0.95}_{-0.95} $ \\
PSZ2 G266.04-21.25 &                & $0.296    $  & $104.6277    $  & $-55.9434    $& $13.92^{+0.29}_{-0.28}                   $& $ 8.72^{+0.06}_{-0.06} $& $46.88^{+0.53}_{-0.53} $ \\
PSZ2 G195.60+44.06 &                & $0.298    $  & $140.1018    $  & $ 30.5028    $& $ 5.36^{+0.09}_{-0.09}                   $& $ 2.63^{+0.02}_{-0.02} $& $ 9.46^{+0.11}_{-0.11} $ \\
PSZ2 G125.71+53.86 &               & $0.302    $  & $189.2441    $  & $ 63.1871    $& $ 5.67^{+0.44}_{-0.39}                    $& $ 2.99^{+0.09}_{-0.09} $& $11.43^{+0.53}_{-0.53} $ \\
PSZ2 G008.94-81.22 &                & $0.307    $  & $  3.5775    $  & $-30.3863    $& $ 9.81^{+0.24}_{-0.24}                   $& $ 6.70^{+0.05}_{-0.05} $& $27.83^{+0.32}_{-0.32} $ \\
PSZ2 G278.58+39.16 &                & $0.308    $  & $172.9775    $  & $-19.9285    $& $ 8.61^{+0.33}_{-0.30}                   $& $ 4.68^{+0.05}_{-0.05} $& $18.93^{+0.31}_{-0.31} $ \\
PSZ1 G103.58+24.78 &  & $0.330    $  & $286.1007    $  & $ 72.4622    $& $ 4.89^{+0.24}_{-0.23}                                 $& $ 1.45^{+0.03}_{-0.03} $& $ 6.09^{+0.15}_{-0.15} $ \\
PSZ2 G349.46-59.95 &                  & $0.347    $  & $342.1824    $  & $-44.5305    $& $12.66^{+0.27}_{-0.27}                 $& $ 8.49^{+0.06}_{-0.06} $& $44.19^{+0.63}_{-0.63} $ \\
PSZ2 G083.29-31.03 &                   & $0.412    $  & $337.1405    $  & $ 20.6204    $& $ 8.70^{+0.28}_{-0.28}                $& $ 5.24^{+0.05}_{-0.05} $& $23.01^{+0.43}_{-0.43} $ \\
PSZ2 G284.41+52.45 &                   & $0.441    $  & $181.5521    $  & $ -8.8002    $& $10.76^{+0.30}_{-0.30}                $& $ 6.79^{+0.06}_{-0.06} $& $33.82^{+0.65}_{-0.65} $ \\
PSZ2 G056.93-55.08 &                    & $0.444    $  & $340.8387    $  & $ -9.5947    $& $ 9.10^{+0.12}_{-0.12}               $& $ 7.30^{+0.03}_{-0.03} $& $31.65^{+0.26}_{-0.26} $ \\
PSZ2\, G254.08-58.45 &                   & $0.458    $  & $ 46.0705    $  & $-44.0257    $& $ 5.84^{+0.33}_{-0.33}              $& $ 3.19^{+0.06}_{-0.06} $& $11.74^{+0.32}_{-0.32} $ \\
PSZ2 G265.10-59.50 &                     & $0.500    $  & $ 40.9129    $  & $-48.5611    $& $ 6.13^{+0.63}_{-0.45}              $& $ 3.10^{+0.08}_{-0.08} $& $12.46^{+0.57}_{-0.57} $ \\
PSZ2 G044.77-51.30 &                     & $0.503    $  & $333.7383    $  & $-14.0045    $& $ 7.95^{+0.44}_{-0.43}              $& $ 5.29^{+0.08}_{-0.08} $& $22.64^{+0.66}_{-0.66} $ \\
PSZ2 G211.21+38.66 &                     & $0.505    $  & $137.7970    $  & $ 17.7760    $& $ 5.48^{+0.22}_{-0.22}              $& $ 3.03^{+0.04}_{-0.04} $& $11.51^{+0.26}_{-0.26} $ \\
 PSZ2 G004.45-19.55 &                      & $0.516    $  & $289.2692    $  & $-33.5228    $& $ 8.73^{+0.48}_{-0.48}            $& $ 6.28^{+0.15}_{-0.15} $& $28.93^{+1.05}_{-1.05} $ \\
PSZ2 G110.28-87.48 &                        & $0.520    $  & $ 12.2939    $  & $-24.6792    $& $ 4.83^{+0.46}_{-0.36}           $& $ 2.30^{+0.08}_{-0.08} $& $ 8.25^{+0.38}_{-0.38} $ \\
PSZ2 G212.44+63.19 &                       & $0.529    $  & $163.2159    $  & $ 24.2584    $& $ 4.15^{+0.23}_{-0.23}            $& $ 1.93^{+0.04}_{-0.04} $& $ 7.08^{+0.25}_{-0.25} $ \\
PSZ2 G201.50-27.31 &                         & $0.538    $  & $ 73.5471    $  & $ -3.0162    $& $ 7.90^{+0.30}_{-0.29}          $& $ 6.47^{+0.07}_{-0.07} $& $28.31^{+0.57}_{-0.57} $ \\
PSZ2 G094.56+51.03 &                        & $0.539    $  & $227.0821    $  & $ 57.9164    $& $ 6.15^{+0.25}_{-0.24}           $& $ 3.87^{+0.05}_{-0.05} $& $14.86^{+0.34}_{-0.34} $ \\
PSZ2 G228.16+75.20 &                        & $0.544    $  & $177.3976    $  & $ 22.4011    $& $ 9.36^{+0.64}_{-0.62}           $& $ 7.08^{+0.17}_{-0.17} $& $31.46^{+1.15}_{-1.15} $ \\
PSZ2 G180.25+21.03 &                         & $0.546    $  & $109.3800    $  & $ 37.7587    $& $12.83^{+0.17}_{-0.17}          $& $11.41^{+0.04}_{-0.04} $& $56.07^{+0.41}_{-0.41} $ \\
 PSZ2 G111.61-45.71 &                           & $0.546    $  & $  4.6399    $  & $ 16.4362    $& $ 9.21^{+0.24}_{-0.24}       $& $ 7.49^{+0.07}_{-0.07} $& $34.24^{+0.52}_{-0.52} $ \\
PSZ2 G183.90+42.99 &                           & $0.559    $  & $137.7032    $  & $ 38.8357    $& $ 8.44^{+0.60}_{-0.53}        $& $ 4.79^{+0.10}_{-0.10} $& $22.54^{+0.78}_{-0.78} $ \\
PSZ2 G155.27-68.42 &                             & $0.567    $  & $ 24.3536    $  & $ -8.4557    $& $ 8.01^{+0.46}_{-0.38}      $& $ 4.20^{+0.06}_{-0.06} $& $18.41^{+0.51}_{-0.51} $ \\
PSZ2 G046.13+30.72 &                             & $0.569    $  & $259.2742    $  & $ 24.0737    $& $ 3.17^{+0.22}_{-0.22}      $& $ 1.26^{+0.04}_{-0.04} $& $ 4.40^{+0.20}_{-0.20} $ \\
PSZ2 G239.93-39.97 &                               & $0.580    $  & $ 71.6966    $  & $-37.0625    $& $ 5.73^{+0.26}_{-0.23}    $& $ 3.36^{+0.04}_{-0.04} $& $13.89^{+0.31}_{-0.31} $ \\
 PSZ2 G254.64-45.20 &                              & $0.581    $  & $ 64.3464    $  & $-47.8134    $& $ 4.97^{+0.27}_{-0.27}    $& $ 3.02^{+0.06}_{-0.06} $& $11.69^{+0.40}_{-0.40} $ \\
PSZ2 G144.83+25.11 &                               & $0.584    $  & $101.9590    $  & $ 70.2481    $& $ 7.78^{+0.21}_{-0.20}    $& $ 4.80^{+0.04}_{-0.04} $& $22.72^{+0.34}_{-0.34} $ \\
 PSZ2 G045.32-38.46 &                               & $0.589    $  & $322.3591    $  & $ -7.6913    $& $ 7.39^{+0.66}_{-0.65}   $& $ 4.95^{+0.16}_{-0.16} $& $22.33^{+1.18}_{-1.18} $ \\
PSZ2 G070.89+49.26 &                               &  $0.602    $  & $239.1098    $  & $ 44.6772    $& $ 5.02^{+0.20}_{-0.21}   $& $ 2.93^{+0.05}_{-0.05} $& $11.71^{+0.31}_{-0.31} $ \\
PSZ2 G045.87+57.70 &                                & $0.609    $  & $229.5866    $  & $ 29.4603    $& $ 5.82^{+0.22}_{-0.22}   $& $ 5.07^{+0.06}_{-0.06} $& $19.56^{+0.35}_{-0.35} $ \\
PSZ2 G073.31+67.52 &                                & $0.609    $  & $215.1709    $  & $ 39.9187    $& $ 6.15^{+0.26}_{-0.25}   $& $ 4.34^{+0.07}_{-0.07} $& $17.97^{+0.47}_{-0.47} $ \\
PSZ2 G099.86+58.45 &                                & $0.615    $  & $213.6952    $  & $ 54.7840    $& $ 7.09^{+0.42}_{-0.42}   $& $ 5.01^{+0.09}_{-0.09} $& $22.03^{+0.79}_{-0.79} $ \\
 PSZ2 G193.31-46.13 &                                & $0.620    $  & $ 53.9644    $  & $ -6.9758    $& $ 5.49^{+0.30}_{-0.32}  $& $ 3.25^{+0.04}_{-0.04} $& $12.88^{+0.26}_{-0.26} $ \\
PLCK\,G147.32-16.59   &                              & $0.645    $  & $ 44.1056    $  & $ 40.2885    $& $ 6.51^{+0.29}_{-0.28}  $& $ 4.26^{+0.05}_{-0.05} $& $17.69^{+0.38}_{-0.38} $ \\
     PLCK\,G$260.7-26.3$  &                         & $0.680    $  & $ 94.1429    $  & $-52.4518    $& $ 4.56^{+0.34}_{-0.32}   $& $ 2.86^{+0.06}_{-0.06} $& $10.57^{+0.35}_{-0.35} $ \\
   PSZ2\,G$097.52+51.70$  &                        &$0.700    $  & $223.8379    $  & $ 58.8718    $& $ 4.08^{+0.23}_{-0.23}     $& $ 3.17^{+0.07}_{-0.07} $& $10.63^{+0.39}_{-0.39} $ \\
PSZ2\,G219.89-34.39 &                              &  $0.700    $  & $ 73.6894    $  & $-20.2851    $& $ 6.77^{+0.33}_{-0.29}   $& $ 4.04^{+0.06}_{-0.06} $& $19.31^{+0.53}_{-0.53} $ \\
   PSZ1\,G$080.66-57.87$  &                      &$0.700    $  & $351.8654    $  & $ -2.0771    $& $ 8.49^{+0.70}_{-0.55}       $& $10.56^{+0.22}_{-0.22} $& $43.59^{+1.53}_{-1.53} $ \\
PSZ2 G208.61-74.39 &                             & $0.711    $  & $ 30.0695    $  & $-24.9132    $& $ 5.23^{+0.23}_{-0.23}      $& $ 3.83^{+0.06}_{-0.06} $& $15.88^{+0.37}_{-0.37} $ \\
   PSZ1\,G$226.65+28.43$  &                   & $0.724    $  & $134.0858    $  & $  1.7803    $& $ 3.44^{+0.20}_{-0.20}         $& $ 2.55^{+0.04}_{-0.04} $& $ 8.39^{+0.25}_{-0.25} $ \\
   PSZ2\,G$084.10+58.72$  &                   & $0.731    $  & $222.2535    $  & $ 48.5569    $& $ 2.93^{+0.17}_{-0.18}         $& $ 1.84^{+0.03}_{-0.03} $& $ 5.65^{+0.16}_{-0.16} $ \\
   PSZ2\,G$087.39+50.92$  &                   & $0.748    $  & $231.6379    $  & $ 54.1523    $& $ 3.29^{+0.30}_{-0.29}         $& $ 2.41^{+0.07}_{-0.07} $& $ 7.55^{+0.34}_{-0.34} $ \\
   PSZ2\,G$088.98+55.07$  &                   & $0.754    $  & $224.7448    $  & $ 52.8167    $& $ 1.46^{+0.23}_{-0.20}         $& $ 0.54^{+0.03}_{-0.03} $& $ 1.57^{+0.12}_{-0.12} $ \\
   PSZ2\,G$086.93+53.18$  &                   & $0.771    $  & $228.5022    $  & $ 52.8035    $& $ 5.21^{+0.19}_{-0.19}         $& $ 4.63^{+0.06}_{-0.06} $& $17.55^{+0.42}_{-0.42} $ \\
   PLCK\,G$079.95+46.96$  &                   & $0.790    $  & $240.5461    $  & $ 51.0615    $& $ 1.99^{+0.23}_{-0.20}         $& $ 0.98^{+0.03}_{-0.03} $& $ 2.64^{+0.11}_{-0.11} $ \\
PSZ2 G352.05-24.01 &                              & $0.798    $  & $290.2490    $  & $-45.8500    $& $ 5.50^{+0.32}_{-0.31}     $& $ 4.40^{+0.07}_{-0.07} $& $20.12^{+0.51}_{-0.51} $ \\
   PLCK\,G$227.99+38.11$  &                     & $0.810    $  & $143.0891    $  & $  5.6900    $& $ 2.55^{+0.59}_{-0.41}       $& $ 1.15^{+0.07}_{-0.07} $& $ 4.15^{+0.48}_{-0.48} $ \\
PSZ2\,G091.83+26.11 &                               & $0.816    $  & $277.7924    $  & $ 62.2429    $& $ 9.83^{+0.52}_{-0.50}   $& $10.62^{+0.20}_{-0.20} $& $53.33^{+1.87}_{-1.87} $ \\
   PSZ2\,G$208.57-44.31$  &                        & $0.830    $  & $ 60.6477    $  & $-15.6784    $& $ 0.65^{+0.22}_{-0.14}    $& $ 0.35^{+0.04}_{-0.04} $& $ 0.66^{+0.07}_{-0.07} $ \\
   PSZ2\,G$071.82-56.55$  &                          & $0.870    $  & $347.3914    $  & $ -4.1707    $& $ 4.42^{+0.27}_{-0.21}  $& $ 4.96^{+0.09}_{-0.09} $& $17.64^{+0.50}_{-0.50} $ \\
   PSZ2\,G$160.83+81.66$  &                         & $0.890    $  & $186.7420    $  & $ 33.5463    $& $ 5.95^{+0.23}_{-0.22}   $& $ 5.06^{+0.07}_{-0.07} $& $22.03^{+0.49}_{-0.49} $ \\
   PSZ1\,G$254.58-32.16$  &                          & $0.900    $  & $ 83.9566    $  & $-48.0247    $& $ 3.80^{+0.31}_{-0.31}  $& $ 2.38^{+0.06}_{-0.06} $& $ 8.59^{+0.32}_{-0.32} $ \\
    SPT-CL\,J$2146-4633$  &                           & $0.933    $  & $326.6447    $  & $-46.5475    $& $ 3.24^{+0.15}_{-0.16} $& $ 2.98^{+0.04}_{-0.04} $& $ 9.75^{+0.23}_{-0.23} $ \\
     PLCK\,G$266.6-27.3$  &                            & $0.972    $  & $ 93.9660    $  & $-57.7796    $& $ 8.93^{+0.40}_{-0.40}$& $10.27^{+0.18}_{-0.18} $& $54.62^{+1.65}_{-1.65} $ \\
    SPT-CL\,J$2341-5119$  &                           & $1.003    $  & $355.3010    $  & $-51.3286    $& $ 4.17^{+0.21}_{-0.20} $& $ 3.68^{+0.07}_{-0.07} $& $15.06^{+0.43}_{-0.43} $ \\
    SPT-CL\,J$0546-5345$  &                           & $1.066    $  & $ 86.6551    $  & $-53.7596    $& $ 4.33^{+0.21}_{-0.19} $& $ 3.67^{+0.05}_{-0.05} $& $15.25^{+0.32}_{-0.32} $ \\
    SPT-CL\,J$2106-5844$  &                            & $1.132    $  & $316.5221    $  & $-58.7421    $& $ 7.01^{+0.49}_{-0.48}$& $ 9.33^{+0.23}_{-0.23} $& $47.17^{+1.95}_{-1.95} $ \\

\end{longtable}
}

\begin{appendix}


\section{Sample data}

Tables~\ref{tab:obsdetails1}, \ref{tab:obsdetails2},  \ref{tab:obsdetails3}, and ~\ref{tab:obsdetails4} contain the sample observation details, including: cluster name(s), redshift, coordinates, column density, exposure time, and \xmm\ OBSID used for the analysis.
%
%
\begin{sidewaystable*}
\centering
\resizebox{1.\textwidth}{!} {
\begin{tabular}{llllcccccr}
\toprule
\toprule
    \planck\ name & SPT name & ACT name & Alt. name & z & RA & Dec. & $N_{\mathrm H}$  & Exp.  & Obs. Id  \\
& &   &   &  &  & &  &  MOS1-2, PN &  \\
 & &   &   &  & [J2000] & [J2000] & [$10^{20} $cm$^{-3}$] &  [ks] &   \\
\midrule

                                    & & & {\bf RXC\,J2023.0 -2056}, AS0868 & 0.056 & 305.7450 & $-20.9485$ & 5.60  & 17, 9 & 201902301\\
                                    & & & {\bf RXC\,J2157.4 -0747}, A2399 & 0.058 & 329.3673 & $-7.8046$ & 3.50 & 10, 7 & 201902801\\
PSZ1\,G246.01-51.76 & & & {\bf RXC\,J0345.7 -4112}, AS0384 & 0.060 & 56.4428 & $-41.2042$ &  1.90 & 17, 8 & 201900801\\
                                    & & & {\bf RXC\,J0225.1 -2928} & 0.060 & 36.2877 & $-29.4773$ & 1.70 & 20, 17 & 302610601 \\
                                    & & & {\bf RXC\,J1236.7 -3354}, AS0700 & 0.080 & 189.1712 & $-33.9260$ & 5.60 & 24, 18 & 302610701 \\
PSZ2\,G346.86-45.38 & & & {\bf RXC\,J2129.8 -5048}, A3771 & 0.080 & 322.4271 & $-50.8167$ & 2.20 & 23, 13 & 201902501\\
PSZ2 G093.92+34.92 &  & &A2255 & $ 0.081 $ & $ 258.2029 $ & $  64.0636 $ & $  2.50 $ & $            6 $, $            4 $ & 011226081 \\ 
PSZ2\,G222.52+20.58 & & & {\bf RXC\,J0821.8 + 0112}, A653 & 0.082 & 125.4614 & 1.1967 & 4.20 & 11, 7 & 201903601\\
PSZ2 G306.77+58.61 &  & &A1651 & $ 0.084 $ & $ 194.8438 $ & $  -4.1983 $ & $  1.81 $ & $            7 $, $            4 $ & 020302011 \\ 
PSZ2 G306.66+61.06 &  & &A1650 & $ 0.084 $ & $ 194.6734 $ & $  -1.7622 $ & $  0.72 $ & $           34 $, $           28 $ & 009320011 \\ 
PSZ2\,G308.64+60.26 & & & {\bf RXC\,J1302.8 -0230}, A1663 &  0.085 & 195.7217 & $-2.51695$ & 1.70 & 26, 16 & 201901801\\
PSZ2\,G099.57-58.64 & & & {\bf RXC\,J0003.8 +0203}, A2700 & 0.092 & 0.9572  & 2.0657 & 3.00 &  26, 19 & 201900101\\
PSZ2 G321.98-47.96 & SPT-CLJ2249-6426 & &A3921 & $ 0.094 $ & $ 342.4917 $ & $ -64.4294 $ & $  1.61 $ & $           30 $, $           23 $ & 011224011 \\ 
PSZ2 G336.60-55.43 &  & &A3911 & $ 0.097 $ & $ 341.5711 $ & $ -52.7261 $ & $  1.50 $ & $           23 $, $           10 $ & 014967031 \\ 
PSZ2\,G311.62-42.31  & & & {\bf RXC\,J2319.6 - 7313}, A3992 & 0.098 & 349.9168 & $-73.2277$ & 1.90 & 10, 6 & 201903301 \\
PSZ2\,G332.23-46.37 & SPT-CLJ2201-5956 & &A3827 & $ 0.099 $ & $ 330.4720 $ & $ -59.9454 $ & $  2.09 $ & $           21 $, $           11 $ & 014967011 \\ 
                                    & & & {\bf RXC\,J0211.4 - 4017}, A2984 & 0.101 & 32.8528 & $-40.2915$ & 1.40 & 29, 22 & 201900601\\
                                    & & & {\bf RXC\,J0049.4 - 2931}, AS0084  & 0.108 & 12.3459 & $-29.5206$ & 1.80 & 20, 13 & 201900401 \\
PSZ2 G053.53+59.52 &  & &A2034 & $ 0.113 $ & $ 227.5528 $ & $  33.5104 $ & $  1.54 $ & $           10 $, $            6 $ & 030393011 \\ 
PSZ2\,G352.28-77.66   & & & {\bf RXC\,J0006.0 - 3443}, A2721 & 0.115 & 1.5015 & $-34.7224$ & 1.20 & 12, 6 & 201903801\\
PSZ2\,G255.64-25.30   & & & {\bf RXC\,J0616.8 -4748} & 0.116 & 94.2158 & $-47.7950$ & 4.80 & 23, 19 & 302610401 \\
PSZ2\,G285.52-62.23   & SPT-CL\,J0145-5301  & ACT-CL\,J0145-5301 & {\bf RXC\,J0145.0 -5300}, A2941 & 0.117 & 26.2433  &  $-53.02085$ & 2.30 & 1,0 & 201900501\\
                                      & & & {\bf RXC,J1516.3 +0005}, A2050 & 0.118 & 229.0747 & 0.0893 & 4.60 & 27, 21 & 201902001\\
                                      & & & {\bf RXC\,J2149.1 -3041}, A3814 & 0.118 & 327.2817 & $-30.7013$ & 2.30 & 25, 18 & 201902601 \\
PSZ2\, G277.38+47.07 & & & {\bf RXC\,J1141.4 -1216}, A1348 & 0.119 & 175.3513  & $-12.2776$ & 3.30 & 28, 22 & 201901601 \\
PSZ2\, G000.04+45.13 & & & {\bf RXC\,J1516.5 -0056}, A2051 & 0.120 & 229.1842  & $-0.9696$ & 5.50 & 29, 22, & 201902101 \\
PSZ2\, G256.38+44.04 & & & {\bf RXC\,J1044.5 -0704}, A1084 & 0.134 & 161.1370  & $-7.0687$ & 3.40 & 26, 18 & 201901501 \\
PSZ2 G241.79-24.01 &  & & {\bf RXC\,J0605.8 -3518}, A3378 & $ 0.139 $ & $  91.4752 $ & $ -35.3023 $ & $  4.02 $ & $           14 $, $           11 $ & 020190101 \\ 
PSZ2\,G042.81-82.97 &  & & {\bf RXC\,J0020.7  -2542}, A0022  & 0.141 & 5.1755 & $-25.7080$ & 2.3 & 15, 11 & 201900301 \\
PSZ2 G002.77-56.16 &  & &{\bf RXC\,J2218.6 -3853}, A3856 & $ 0.141 $ & $ 334.6677 $ & $ -38.9018 $ & $  1.13 $ & $           14 $, $            4 $ & 020190301 \\ 
PSZ2 G226.18+76.79 &  & &A1413 & $ 0.143 $ & $ 178.8250 $ & $  23.4049 $ & $  1.84 $ & $           62 $, $           42 $ & 050269021 \\ 
PSZ2\,G028.77-33.56 & & & {\bf RXC\,J2048.1 $-1750$}, A2328  & 0.147 & 312.0419  & -17.8413 & 4.80 & 25, 19 & 201902401 \\
PSZ2 G236.92-26.65 &  & &{\bf RXC\,J0547.6 -3152}, A3364 & $ 0.148 $ & $  86.9081 $ & $ -31.8727 $ & $  2.07 $ & $           22 $, $           14 $ & 020190091 \\ 
PSZ2 G008.47-56.34 &  & & {\bf RXC\,J2217.7 -3543}, A3854 & $ 0.149 $ & $ 334.4400 $ & $ -35.7260 $ & $  1.20 $ & $           22 $, $           14 $ & 020190291 \\ 
PSZ2 G003.93-59.41 &  & &{\bf RXC\,J2234.5 -3744}, A3888 & $ 0.151 $ & $ 338.6125 $ & $ -37.7360 $ & $  1.32 $ & $           20 $, $           13 $ & 040491081 \\ 
PSZ2 G021.10+33.24 &  & &A2204 & $ 0.152 $ & $ 248.1959 $ & $   5.5754 $ & $  6.97 $ & $           14 $, $            8 $ & 030649021 \\ 
PSZ2 G244.71+32.50 &  & & RXC\,J0945.4 -0839, A0868 & $ 0.153 $ & $ 146.3575 $ & $  -8.6557 $ & $  3.59 $ & $            8 $, $            5 $ & 001754011 \\ 
PSZ2\,G018.32-28.50 &   &  & {\bf RXC J2014.8 -2430} & 0.154 & 303.7154 & $-24.5059$ & 7.40 & 25, 16 &  201902201 \\
PSZ2 G049.22+30.87 &  & &RX\,J1720.1+2638 & $ 0.164 $ & $ 260.0417 $ & $  26.6250 $ & $  5.65 $ & $           14 $, $            8 $ & 050067041 \\ 
PSZ2 G263.68-22.55 & SPT-CLJ0645-5413 & ACT-CL J0645-5413& {\bf RXC\,J0645.4 -5413}, A3404 & $ 0.164 $ & $ 101.3712 $ & $ -54.2273 $ & $  5.60 $ & $           11 $, $            7 $ & 040491041 \\ 
\bottomrule

\end{tabular}
}
    \caption{\footnotesize  Sample observation details. Columns: (1) \planck\ name; (2) SPT name; (3) ACT name (4) alternative name (non-exhaustive), \rexcess\ clusters in bold; (5) redshift; (6) RA; (7) Dec.; (8) neutral hydrogen column density integrated along the line of sight determined from the LAB survey \citep{kal05}; (9) observation exposure time for MOS1 and pn detectors, in ks; (10) \xmm\ OBSID.}\label{tab:obsdetails1}
    \label{tab:obs1}
\end{sidewaystable*}
%
\begin{sidewaystable*}
\centering
\resizebox{1.\textwidth}{!} {
\begin{tabular}{llllcccccr}
\toprule
\toprule
    \planck\ name & SPT name & ACT name & Alt. name & z & RA & Dec. & $N_{\mathrm H}$  & Exp.  & Obs. Id  \\
& &   &   &  &  & &  &  MOS1-2, PN &  \\
 & &   &   &  & [J2000] & [J2000] & [$10^{20} $cm$^{-3}$] &  [ks] &   \\
\hline
\\
PSZ2\, G249.38+33.26 & & & {\bf RXC J0958.3 -1103}, A0907 & 0.167 & 149.5929  & $-11.0644$ & 5.10 & 9, 5 & 201903501 \\
PSZ2 G097.72+38.12 &  & &A2218 & $ 0.171 $ & $ 248.9597 $ & $  66.2125 $ & $  2.60 $ & $           17 $, $           11 $ & 011298011 \\ 
PSZ2 G067.17+67.46 &  & &A1914 & $ 0.171 $ & $ 216.5105 $ & $  37.8243 $ & $  1.06 $ & $           15 $, $            7 $ & 011223021 \\ 
PSZ2 G149.75+34.68 &  & &A0665 & $ 0.182 $ & $ 127.7462 $ & $  65.8398 $ & $  4.24 $ & $            5 $, $            2 $ & 010989041 \\ 
PSZ2 G313.33+61.13 &  & & {\bf RXC\,J1311.4 -0120}, A1689 & $ 0.183 $ & $ 197.8726 $ & $  -1.3417 $ & $  1.52 $ & $           36 $, $           27 $ & 009303011 \\
PSZ2 G195.75-24.32 &  & &A520 & $ 0.203 $ & $  73.5402 $ & $   2.9212 $ & $  3.74 $ & $           20 $, $            9 $ & 020151011 \\ 
PSZ2 G006.76+30.45 &  & &A2163 & $ 0.203 $ & $ 243.9399 $ & $  -6.1491 $ & $ 16.50 $ & $           10 $, $            6 $ & 011223061 \\ 
PSZ2 G182.59+55.83 &  & &A963 & $ 0.206 $ & $ 154.2653 $ & $  39.0482 $ & $  1.25 $ & $           18 $, $           12 $ & 008423071 \\ 
PSZ2 G166.09+43.38 &  & &A773 & $ 0.217 $ & $ 139.4765 $ & $  51.7315 $ & $  1.28 $ & $           13 $, $           14 $ & 008423061 \\ 
PSZ2 G092.71+73.46 &  & &A1763 & $ 0.223 $ & $ 203.8298 $ & $  41.0001 $ & $  0.94 $ & $           12 $, $            9 $ & 008423091 \\ 
PSZ2\,G055.59+31.85 &  & & A2261 & 0.224 & 260.6127 & 32.1324 & 3.19 & 2, 1 & 0093031001 \\ 
PSZ2 G072.62+41.46 &  & &A2219 & $ 0.228 $ & $ 250.0837 $ & $  46.7107 $ & $  1.76 $ & $           12 $, $            6 $ & 060500051 \\ 
PSZ2 G073.97-27.82 &  & &A2390 & $ 0.231 $ & $ 328.4031 $ & $  17.6949 $ & $  8.66 $ & $           10 $, $            8 $ & 011127011 \\ 
PSZ2 G294.68-37.01 &  & &RXCJ0303.8-7752 & $ 0.274 $ & $  45.9366 $ & $ -77.8784 $ & $  8.73 $ & $           11 $, $            8 $ & 020533011 \\ 
PSZ2 G241.76-30.88 &  & &RXCJ0532.9-3701 & $ 0.275 $ & $  83.2323 $ & $ -37.0270 $ & $  2.90 $ & $           11 $, $            6 $ & 004234181 \\ 
PSZ2 G259.98-63.43 & SPT-CLJ0232-4421 & &RXCJ0232.2-4420 & $ 0.284 $ & $  38.0772 $ & $ -44.3464 $ & $  2.49 $ & $           12 $, $            7 $ & 004234031 \\ 
PSZ2 G244.37-32.15 &  & &RXC\,J0528.9-3927 & $ 0.284 $ & $  82.2211 $ & $ -39.4714 $ & $  2.13 $ & $            7 $, $            4 $ & 004234081 \\ 
PSZ2 G106.87-83.23 &  & &RXC\,J0043.4 -2037, A2813 & $ 0.292 $ & $  10.8519 $ & $ -20.6229 $ & $  1.54 $ & $           11 $, $            5 $ & 004234021 \\ 
PSZ2 G262.27-35.38 & SPT-CL J0516-5430 & ACT-CL J0516-5430 & ACO S520 & 0.295 & 79.1536 &  $-54.5089$ & 2.56 & 3, 1 & 0042340701 \\
 PSZ2 G266.04-21.25 & SPT-CLJ0658-5556 & ACT-CL J0658-5557&1ES0657-558 & $ 0.296 $ & $ 104.6277 $ & $ -55.9434 $ & $  4.17 $ & $           21 $, $           14 $ & 011298021 \\ 
PSZ2 G195.60+44.06 &  & & A781 & $ 0.298 $ & $ 140.1018 $ & $  30.5028 $ & $  1.94 $ & $           57 $, $           47 $ & 040117011 \\ 
PSZ2 G125.71+53.86 &  & &A1576 & $ 0.302 $ & $ 189.2441 $ & $  63.1871 $ & $  1.68 $ & $            6 $, $            1 $ & 040225011 \\ 
PSZ2 G008.94-81.22 &  & &A2744 & $ 0.307 $ & $   3.5775 $ & $ -30.3863 $ & $  1.60 $ & $           14 $, $           10 $ & 004234011 \\ 
PSZ2 G278.58+39.16 &  & &A1300 & $ 0.308 $ & $ 172.9775 $ & $ -19.9285 $ & $  4.50 $ & $           11 $, $            9 $ & 004234101 \\ 
PSZ1 G103.58+24.78 &                                  &    &                                         & 0.334 & 286.1007 & 72.4622 & 7.48 & 28, 15 &  0693660501 \\
&&&&&&&&& 0693663101 \\
PSZ2 G349.46-59.95 & SPT-CLJ2248-4431 & &AS\,1063 & $ 0.347 $ & $ 342.1824 $ & $ -44.5305 $ & $  1.84 $ & $           25 $, $           15 $ & 050463011 \\ 
PSZ2 G083.29-31.03 &  & &RXCJ2228+2037 & $ 0.412 $ & $ 337.1405 $ & $  20.6204 $ & $  4.26 $ & $           24 $, $           15 $ & 014789011 \\ 
PSZ2 G284.41+52.45 &  & &MACSJ1206.2-0848 & $ 0.441 $ & $ 181.5521 $ & $  -8.8002 $ & $  4.35 $ & $           30 $, $           21 $ & 050243041 \\ 
PSZ2 G056.93-55.08 &  & &MACSJ2243.3-0935 & $ 0.444 $ & $ 340.8387 $ & $  -9.5947 $ & $  3.11 $ & $          103 $, $           78 $ & 050349021 \\ 
PSZ2\, G254.08-58.45 & SPT-CL\, J0304-4401 &  & SMACS\,J0304.3-4402 & 0.458 & 46.0705 & $-44.0257$ & 1.29 & 17, 12 & 0700182201 \\
PSZ2 G265.10-59.50 & SPT-CLJ0243-4833 &  & RXCJ0243.6-4834& $ 0.500 $ & $  40.9129 $ & $ -48.5611 $ & $  2.15 $ & $           15,            6 $  & 0672090501 \\
&&&&&&&&&0723780801 \\ 
PSZ2 G044.77-51.30 &  &  & MACSJ2214.9-1359& $ 0.503 $ & $ 333.7383 $ & $ -14.0045 $ & $  2.88 $ & $           17,            7 $  & 0693661901 \\ 
PSZ2 G211.21+38.66 &  &  & MACSJ0911.2+1746& $ 0.505 $ & $ 137.7970 $ & $  17.7760 $ & $  3.28 $ & $           33,           25 $  & 0693662501 \\
PSZ2 G004.45-19.55 &  &  & & $ 0.516 $ & $ 289.2692 $ & $ -33.5228 $ & $  5.90 $ & $           14,            5 $  & 0656201001 \\ 
PSZ2 G110.28-87.48 &  &  & & $ 0.520 $ & $  12.2939 $ & $ -24.6792 $ & $  1.50 $ & $           26,            4 $  & 0693662101 \\
&&&&&&&&&0723780201 \\ 
PSZ2 G212.44+63.19 &  &  & RMJ105252.4+241530.0& $ 0.529 $ & $ 163.2159 $ & $  24.2584 $ & $  1.89 $ & $           34,           20 $  & 0693660701 \\
&&&&&&&&& 0723780701 \\ 
\bottomrule
    \end{tabular}
    }
    \caption{\footnotesize Continued.}\label{tab:obsdetails2}
    \label{tab:obs2}
\end{sidewaystable*}
%
\begin{sidewaystable*}
\centering
\resizebox{1.\textwidth}{!} {
\begin{tabular}{llllcccccr}
\toprule
\toprule
    \planck\ name & SPT name & ACT name & Alt. name & z & RA & Dec. & $N_{\mathrm H}$  & Exp.  & Obs. Id  \\
& &   &   &  &  & &  &  MOS1-2, PN &  \\
 & &   &   &  & [J2000] & [J2000] & [$10^{20} $cm$^{-3}$] &  [ks] &   \\
\midrule
PSZ2 G201.50-27.31 &  &  & MACSJ0454.1-0300& $ 0.538 $ & $  73.5471 $ & $  -3.0162 $ & $  3.92 $ & $           22,           17 $  & 0205670101 \\
PSZ2 G094.56+51.03 &  &  & WHL J227.050+57.90& $ 0.539 $ & $ 227.0821 $ & $  57.9164 $ & $  1.50 $ & $           26,           20 $  & 0693660101 \\
&&&&&&&&& 0723780501 \\ 
PSZ2 G228.16+75.20 &  &  & MACSJ1149.5+2223& $ 0.544 $ & $ 177.3976 $ & $  22.4011 $ & $  1.92 $ & $           13,            2 $  & 0693661701 \\ 
PSZ2 G111.61-45.71 &  &  & CL0016+16& $ 0.546 $ & $   4.6399 $ & $  16.4362 $ & $  3.99 $ & $           33,           24 $  & 0111000101 \\
&&&&&&&&&0111000201 \\ 
PSZ2 G180.25+21.03 &  &  & MACSJ0717.5+3745& $ 0.546 $ & $ 109.3800 $ & $  37.7587 $ & $  6.63 $ & $          156,          116 $  & 0672420101 \\
&&&&&&&&&0672420201 \\
&&&&&&&&& 0672420301 \\ 
PSZ2 G183.90+42.99 &  &  & WHL J137.713+38.83& $ 0.559 $ & $ 137.7032 $ & $  38.8357 $ & $  1.63 $ & $           14,            8 $  & 0723780101 \\ 
PSZ2 G155.27-68.42 &  &  & WHL J24.3324-8.477& $ 0.567 $ & $  24.3536 $ & $  -8.4557 $ & $  3.52 $ & $           28,           18 $  & 0693662801\\
&&&&&&&&& 0700180201 \\ 
PSZ2 G046.13+30.72 &  &  & WHL J171705.5+240424& $ 0.569 $ & $ 259.2742 $ & $  24.0737 $ & $  5.18 $ & $           26,            2 $  & 0693661401 \\ 
PSZ2 G239.93-39.97 &  &  & & $ 0.580 $ & $  71.6966 $ & $ -37.0625 $ & $  1.44 $ & $           34,           23 $  & 0679181001 \\
&&&&&&&&& 0693661201 \\ 
PSZ2 G254.64-45.20 & SPT-CLJ0417-4748 &  & & $ 0.581 $ & $  64.3464 $ & $ -47.8134 $ & $  1.34 $ & $           20,            9 $  & 0700182401 \\ 
PSZ2 G144.83+25.11 &  &  & MACSJ20647.7+7015& $ 0.584 $ & $ 101.9590 $ & $  70.2481 $ & $  5.40 $ & $           72,           46 $  & 0551850401 \\
&&&&&&&&&0551851301 \\ 
PSZ2 G045.32-38.46 &  &  & MACSJ2129.4-0741& $ 0.589 $ & $ 322.3591 $ & $  -7.6913 $ & $  4.32 $ & $            9,            3 $  & 0700182001  \\
PSZ2 G070.89+49.26 &  &  & WHL J155625.2+444042& $ 0.602 $ & $ 239.1098 $ & $  44.6772 $ & $  1.23 $ & $           47,           25 $  & 0693661301 \\ 
PSZ2 G045.87+57.70 &  &  & WHL J151820.6+292740& $ 0.609 $ & $ 229.5866 $ & $  29.4603 $ & $  2.12 $ & $           23,           15 $  & 0693661101 \\ 
PSZ2 G073.31+67.52 &  &  & WHL J215.168+39.91& $ 0.609 $ & $ 215.1709 $ & $  39.9187 $ & $  0.82 $ & $           26,           19 $  & 0693661001 \\ 
PSZ2 G099.86+58.45 &  &  & WHL J213.697+54.78& $ 0.615 $ & $ 213.6952 $ & $  54.7840 $ & $  1.50 $ & $           21,            9 $  & 0693660601 \\
&&&&&&&&&0693662701 \\
&&&&&&&&&0723780301 \\ 
PSZ2 G193.31-46.13 &  &  & & $ 0.634 $ & $  53.9644 $ & $  -6.9758 $ & $  4.15 $ & $           67,           48 $  & 0658200401 \\
&&&&&&&&&0693661501 \\ 
PLCK G147.32-16.59  &  &  & & $ 0.645 $ & $  44.1056 $ & $  40.2885 $ & $  8.29 $ & $           59,           39 $  & 0679181301 \\
&&&&&&&&&0693661601 \\ 
PLCK G260.7-26.3 & SPT-CLJ0616-5227 & ACT-CL J0616-5227 & & $ 0.680 $ & $  94.1429 $ & $ -52.4518 $ & $  4.25 $ & $           35,           20 $  & 0693662301 \\ 
%
%
PSZ2\,G097.52+51.70 &                                  &    &                                        & 0.700 & 223.8379 & 58.8718 & 1.06 & 24, 17 & 0783881301 \\
PSZ2\,G219.89-34.39 &  &  & & $ 0.700 $ & $  73.6894 $ & $ -20.2851 $ & $  3.25 $ & $           51,           34 $  & 0679180501 \\
&&&&&&&&& 0693660301 \\ 
PSZ1\,G080.66-57.87  &   & ACT-CL\,J2327.4-0204&  RCS2\,J2327-0204     & 0.699 & 351.8654 & $-2.0771$ & 4.62 & 25, 75, 71 & 7355 \\
&&&&&&&&&14025 \\
&&&&&&&&& 14361 \\
PSZ2 G208.61-74.39 &  &  & & $ 0.711 $ & $  30.0695 $ & $ -24.9132 $ & $  1.37 $ & $           47,           28 $  & 0693662901 \\
&&&&&&&&&0723780601 \\
PSZ1\,G226.65+28.43 &                                  &    &                                         & 0.724 & 134.0858 & 1.7803 & 3.38 & 44, 27 & 0783881001 \\
PSZ2\,G084.10+58.72 &                                  &    &                                         & 0.731 & 222.2535 & 48.5569 & 2.29 & 72, 58 & 0783880901 \\

\bottomrule

    \end{tabular}
    }
    \caption{\footnotesize Continued.}\label{tab:obsdetails3}
    \label{tab:obs3}
\end{sidewaystable*}
%
\begin{sidewaystable*}
    \centering
\resizebox{1.\textwidth}{!} {
\begin{tabular}{llllcccccr}
\toprule
\toprule
    \planck\ name & SPT name & ACT name & Alt. name & z & RA & Dec. & $N_{\mathrm H}$  & Exp.  & Obs. Id  \\
& &   &   &  &  & &  &  MOS1-2, PN &  \\
 & &   &   &  & [J2000] & [J2000] & [$10^{20} $cm$^{-3}$] &  [ks] &   \\
\midrule
PSZ2\,G087.39+50.92 &                                  &    &                                        & 0.748 & 231.6379 & 54.1524 & 1.21 & 20, 11 & 0783881201 \\
PSZ2\,G088.98+55.07  &                                 &   &                                          & 0.754 & 224.7448 & 52.8167 & 1.83 & 44, 30 & 0783880801 \\
PSZ2\,G086.93+53.18 \\ &                               &  &                                           & 0.771 & 228.5022 & 52.8035 & 1.62 & 57, 37 & 0783880701 \\
PLCK\, G079.95+46.96 &                                 &  &                                          & 0.790 & 240.5461 & 51.0615 & 1.56 & 81, 51 & 0783880601 \\
PSZ2 G352.05-24.01 &  &  & & $ 0.798 $ & $ 290.2490 $ & $ -45.8500 $ & $  6.03 $ & $           52,           28 $  & 0679180201 \\
&&&&&&&&&0693660401 \\
PLCK\,G227.99+38.11 &                                  &   &                                         & 0.810 & 143.0891 & 5.6899 & 3.40 & 26, 10 & 0783880501 \\
PSZ2\,G091.83+26.11 &                                  &   &                                         & 0.822 & 277.7925 & 62.2429 &  4.30 & 23, 13 & 0762801001 \\
PSZ2G\,208.57-44.31 &                                  &   &                                          & 0.830 &  60.6477 & $-15.6783$ & 2.13 & 60, 47 & 0783880301 \\
PSZ2\,G071.82-56.55 &                                  &   &                                          & 0.870 & 347.3914 & $-4.1707$ & 3.59 & 46, 27 & 0783880201 \\
PSZ2\,G160.83+81.66 &                                 &  & WARPS J\,1226.9+3332  & 0.892 & 186.7420 &  33.5463 & 1.83 & 63, 47 & 0200340101 \\
PSZ1\,G254.58-32.16 & SPT-CL J0535-4801 &  &                                          & 0.900 & 83.9566   &  -48.0247 & 3.11 & 56, 30 &  0783880101 \\
%
%
 & SPT-CLJ2146-4633 &  & & $ 0.933 $ & $ 326.6447 $ & $ -46.5475 $ & $  1.64 $ & $          153,          102 $  & 0744400501 \\
 &&&&&&&&&0744401301 \\ 
PSZ2 G266.54-27.31 & SPT-CLJ0615-5746 &  & & $ 0.972 $ & $  93.9660 $ & $ -57.7796 $ & $  4.32 $ & $           12,            3 $  & 0658200101 \\ 
 & SPT-CLJ2341-5119 &  & & $ 1.003 $ & $ 355.3010 $ & $ -51.3286 $ & $  1.21 $ & $           91,           45 $  & 0744400401 \\
 &&&&&&&&&0763670201 \\ 
 & SPT-CLJ0546-5345 & ACT-CL J0546-5345 & & $ 1.066 $ & $  86.6551 $ & $ -53.7596 $ & $  6.79 $ & $          127,          113 $  & 0744400201 \\
 &&&&&&&&& 0744400301 \\ 
 & SPT-CLJ2106-5844 &  & & $ 1.132 $ & $ 316.5221 $ & $ -58.7421 $ & $  4.33 $ & $           26,           16 $  & 0763670301 \\ 
\bottomrule
    \end{tabular}
    }
    \caption{\footnotesize Continued.}\label{tab:obsdetails4}
    \label{tab:obs4}
\end{sidewaystable*}


\section{Sample selection bias and covariance tests}\label{appx:sims}

We used similar simulations as those produced for the \citet{fas21} study. Simulated clusters, modelled with the \citet{arn10} pressure profile and drawn from a Tinker mass function \citep{tin08}, were injected into the Planck Early SZ maps in the `cosmological' mask region. The $Y_{\rm SZ}$ value for each object was drawn from the $Y_{\rm SZ}-M$ relation of \citet{arn10}, with a bias between the X-ray calibrated mass and the true mass of $(1-b) =0.65$, a value used to obtain the observed cluster number counts in the \planck\ cosmology. As the slope of the $Y_{\rm SZ} - Y_{\rm X}$ relation is expected to be close to unity, and we are only interested in slope variations, we assume that the normalisation and slope of the $Y_{\rm X}-M$ and $Y_{\rm SZ}-M$ relations are the same. 

We draw the $Y_{\rm SZ}$, $Y_{\rm X}$ and $L_{\rm X}$ quantities associated to each simulated cluster following a correlated Gaussian distribution with covariance matrix. If $\mathcal{Q}_{M}$ is the latent value of $\mathcal{Q}$ at a given mass, obtained if there were no scatter, one can write
\begin{gather}
 Y_{\rm M} = A_{\rm Y} E(z)^{2/3} M^{B_{\rm Y}}\\
    L_{\rm M} =  A_{\rm L} E(z)^2 M^{B_{\rm L}}\\
    P(D_{\rm A}^2 Y_{\rm SZ}, Y_{\rm X}, L_{\rm Xc} | Y_{\rm M}, Y_{\rm M}, L_{\rm M})=\mathcal{N}[(Y_{\rm M}, Y_{\rm M}, L_{\rm M}),V_\sigma],
\end{gather}
where $Y_{\rm M}$ and $L_{\rm M}$ are the latent values of $Y$ and $L$ at a given mass (the values obtained if there were no scatter), and $D_{\rm A}^2 Y_{\rm SZ}$, $Y_{\rm X}$, and $L_{\rm Xc}$ are the true values.  For these simulations we assume $B_{\rm Y} =  1.79$ \citep{arn10} and $B_{\rm L} = 1.37$ (Table~\ref{tab:lmrels}). $\mathcal{N}$ is a Gaussian log-normal correlated distribution at fixed mass, where
\begin{gather}
V_{\sigma} = \
\begin{pmatrix}
\sigma^2_{\ln{Y_{\rm SZ}}} & r \, \sigma_{\ln{Y_{\rm SZ}}} \sigma_{\ln{Y_{\rm X}}} & s \, \sigma_{\ln{Y_{\rm SZ}}} \sigma_{\ln{L_{\rm Xc}}}\\
r \, \sigma_{\ln{Y_{\rm SZ}}} \sigma_{\ln{Y_{\rm X}}} & \sigma^2_{\ln{Y_{\rm X}}} & t \, \sigma_{\ln{Y_{\rm X}}} \sigma_{\ln{L_{\rm Xc}}}\\
s \, \sigma_{\ln{Y_{\rm SZ}}} \sigma_{\ln{L_{\rm Xc}}} & t \, \sigma_{\ln{Y_{\rm X}}} \sigma_{\ln{L_{\rm Xc}}} & \sigma^2_{\ln{L_{\rm Xc}}}
\end{pmatrix}.
\end{gather}
For the covariance matrix $V_\sigma$, we assume $\sigma_{\ln{Y_{\rm SZ}}}=0.12$ \citep{stk12,alb17}; $\sigma_{\ln{Y_{\rm X}}}=0.16$ \citep{sp14,alb17,nt18}; $r=0.4$ \citep{af19,an19}; $s=0.4$ \citep{af19}; and $t=0.85$ \citep{af19}. As detailed in Sect.~\ref{sec:seleff}, with these assumptions, Eqn.~\ref{eqn:forder} gives a first-order estimate of $\sigma_{\ln{L_{\rm Xc}}} = 0.22$ in the absence of selection effects.  

\begin{figure}[!t]
\begin{center}
\includegraphics[scale=1.,angle=0,keepaspectratio,width=0.975\columnwidth]{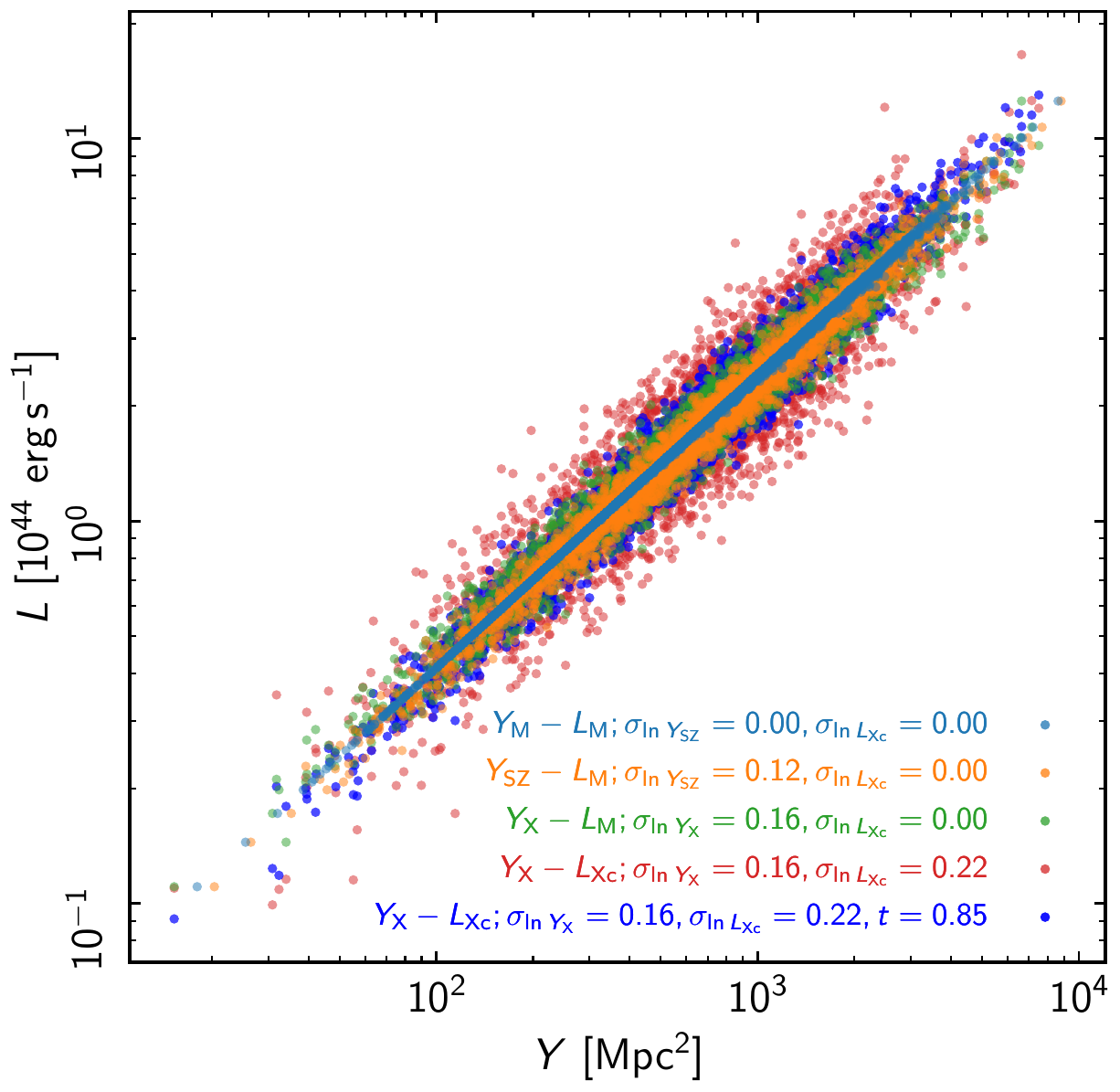}
\end{center}
\caption{\footnotesize  Simulated $Y_{\rm SZ}$ reextracted from PSZ2 maps, plotted as a function of \Lxc. The quantity $\mathcal{Q}_{\rm M}$ is the latent variable with respect to the mass $M$, obtained if there were no scatter in the relation. The Figure shows the effect of progressively adding scatter in $Y_{\rm SZ}, Y_{\rm X}$, and \Lxc\ with $M$. The blue points include covariance of $t=0.85$ between $Y_{\rm X}$ and \Lxc.
}
\label{fig:sims}
\end{figure}

The Multi-Matched Filter extraction algorithm \citep{mel06} was then applied, to obtain SZ detections at ${\rm S/N} > 6$, corresponding to the threshold for the ESZ sample.  We then matched the injected and recovered clusters to produce a mock ESZ catalogue, doing this twenty times, resulting in a total of 3188 detections. Measurement errors were estimated from the maps as described in \citet{mel06}. Fits to the simulated data were performed as described in Sect.~\ref{sec:lxmres}. Our baseline simulation in the following assumes $r=s=t=0$ (i.e. zero covariance between quantities). 

We first fitted the $Y_{\rm M} - L_{\rm M}$ relation, which assumes that there is zero intrinsic scatter in either of the observables with respect to the mass. The scatter in the $Y$-axis seen in Fig.~\ref{fig:sims} is then entirely due to the observational uncertainties in the SZE measurements. The resulting slope of $0.759\pm0.001$ implies a change in the \Lxc$-M$ relation of $\Delta\, B_{\rm L} = -0.01$.  Adding intrinsic scatter of $\sigma_{\ln{Y_{\rm SZ}}}=0.12$ \citep{stk12,alb17}, the slope of the  $Y_{\rm SZ} - L_{\rm M}$ relation is $0.759\pm0.001$, again implying a negligible change of $\Delta\, B_{\rm L} = -0.01$ on the slope of the \Lxc$-M$ relation. These results imply that Malmquist bias in the $Y_{\rm SZ}$ observable are negligible, and cannot account for the significantly steeper than self-similar slope we find for the \Lxc$-M$ relation.

We next studied the robustness of the recovery of the \Lxc$-M$ relation slope to intrinsic scatter in the $Y_{\rm X}$ proxy, due to $Y_{\rm X}$ being a scattered estimates of $Y_{\rm SZ}$. An intrinsic scatter of $\sigma_{\ln{Y_{\rm X}}}=0.16$ \citep{sp14,alb17,nt18} pushes the slope of the $Y_{\rm X} - L_{\rm M}$ relation to $0.741\pm0.002$, in turn changing the  \Lxc$-M$ relation slope by $\Delta\,B_{\rm L} = -0.04$. 

We then added an intrinsic scatter of $\sigma_{\ln{L_{\rm Xc}}} = 0.22$, finding that this redresses the slope to $0.775\pm0.005$, implying a change of $\Delta\,B_{\rm L} = +0.02$ in the slope of the \Lxc$-M$ relation. Finally, we added a covariance of $t=0.85$ between $Y_{\rm X}$ and \Lxc\ \citep{af19}. This pushes the slope to a slightly steeper value of $0.785\pm0.003$, changing the  \Lxc$-M$ relation slope by $\Delta\,B_{\rm L} = +0.04$. The covariance significantly reduces the dispersion in the $Y_{\rm X}-$\Lxc\ relation, by a factor of two. This can clearly be seen in the difference in dispersion between the red and blue points in Fig.~\ref{fig:sims}.

In conclusion, we find that selection effects and intrinsic scatter have a negligible effect on the slope of the \Lxc$-M$ relation, and cannot account for the significantly steeper than self-similar value that we find in this work. The dispersion of the \Lxc$-M$ relation that we derive is significantly underestimated, most likely by a factor of $\sim 1.7$, due to the covariance between \Lxc\ and the mass proxy, $Y_{\rm X}$.

\end{appendix}

\raggedright
\end{document}